# Contactless Pulse Rate Assessment: Results and Insights for Application in Driving Simulator

**Đorđe D. Nešković[1,2], Kristina Stojmenova Pečečnik[3], Jaka Sodnik[3*], Nadica Miljković[1,3]**

[1] University of Belgrade - School of Electrical Engineering, Bulevar kralja Aleksandra 73, 11000 Belgrade, Serbia
[2] Vinča Institute of Nuclear Sciences - National Institute of the Republic of Serbia, University of Belgrade, Mike Petrovića Alasa 12-14, 11351 Vinča, Belgrade, Serbia
[3] Faculty of Electrical Engineering, University of Ljubljana, Tržaška c. 25, 1000 Ljubljana, Slovenia

**Corresponding Author*
*e-mails: djordje.neskovic@vin.bg.ac.rs, kristina.stojmenova@fe.uni-lj.si, jaka.sodnik@fe.uni-lj.si, nadica.miljkovic@etf.bg.ac.rs*

**Abstract**:
Remote photoplethysmography (rPPG) offers a promising solution for non-contact driver monitoring by detecting subtle blood flow-induced facial color changes from video. However, motion artifacts in dynamic driving environments remain key challenges. This study presents an rPPG framework that combines signal processing techniques before and after applying Eulerian Video Magnification (EVM) for pulse rate (PR) estimation in driving simulators. While not novel, the approach offers insights into the efficiency of the EVM method and its time complexity. We compare results of the proposed rPPG approach against reference Empatica E4 data and also compare it with existing achievements from the literature. Additionally, the possible bias of the Empatica E4 is further assessed using an independent dataset with both the Empatica E4 and the Faros 360 measurements. EVM slightly improves PR estimation, reducing the mean absolute error (MAE) from 6.48 bpm to 5.04 bpm (the lowest MAE (~2 bpm) was achieved under strict conditions) with an additional time required for EVM of about 20 s for 30 s sequence. Furthermore, statistically significant differences are identified between younger and older drivers in both reference and rPPG data. Our findings demonstrate the feasibility of using rPPG-based PR monitoring, encouraging further research in driving simulations.

## 1. Introduction

The ability to monitor physiological parameters, such as pulse rate (PR) or the inter beat interval (IBI), offers valuable insights into a driver's physical and mental states, which can directly impact road safety. For example, an elevated PR may indicate stress or fatigue, which may result in risk of accidents. By introducing early detection of physiological changes, specifically PR changes, danger to passengers and other traffic participants can be avoided by warning the driver of exhaustion or by automatically stopping the vehicle in an emergency. Furthermore, in the case of a disaster, systems for contactless monitoring of vital parameters can provide essential information to the emergency services or the police during the investigation [1–3].

In the driving simulation environment, which can be used to assess driving performance [4], wearable sensors based on photoplethysmography (PPG) are commonly employed to assess PR and heart rate variability (HRV) features [5]. PPG sensors operate by emitting light beams and measuring changes in blood volume within tissues, thus providing an indirect estimation of cardiovascular state [6]. PPG sensors are commonly integrated into wearable devices, such as smartwatches and wristbands, including Apple Watch, Fitbit, Garmin, Empatica E4, and Biovotion [7]. While wearable devices offer convenient and continuous monitoring, their accuracy and reliability can be affected by multiple factors, such as motion artifacts and ambient light interference. In contrast, contact-based PPG sensors, commonly attached to the earlobe or incorporated within finger pulse oximeters, provide more reliable and high-fidelity physiological signal measurements compared to wearable bracelets [6]. These devices are often used in clinical and research applications due to their higher accuracy and consistency. Unfortunately, laboratory-graded pulse oximeters and earlobe PPG devices are not suitable for applications in dynamic driving simulation conditions due to their sensitivity to movement, non-compact dimensions, the need for stable positioning, and often wired connections that limit freedom of movement [8].

An alternative approach to PPG sensors can be remote photoplethysmography (rPPG), which is based on video processing technologies, where available research results indicate that subtle changes in skin color caused by blood pulsation can be detected using affordable red, green, and blue (RGB) cameras [9–12]. The use of cameras for detecting physiological parameters offers advantages over wearable devices (e.g., smart-watches, wristbands, or PPG sensors placed on the finger or on the ears). Cameras enable completely unobtrusive, non-invasive, and comfortable monitoring, eliminating the need for physical contact that may cause skin irritation or discomfort during prolonged use [1]. In addition, rPPG technology further reduces user distraction during activities like driving and allows seamless monitoring without additional adjustments or interaction. Moreover, cameras can be easily integrated into existing systems, such as in-car infotainment systems, reducing costs and implementation complexity. Additionally, cameras enable simultaneous monitoring of multiple individuals in a single environment, which wearable devices cannot provide. Another important advantage of camera-based methods is that they rely on light reflected from the skin surface, allowing measurements to be performed from virtually any visible skin region [6]. This stands in contrast to contact-based PPG devices that operate in transmission mode, which are limited to tissue sites where light can pass through, such as the fingertip or earlobe [6]. Further advantages are that cameras require minimal maintenance, eliminate the need for direct contact with the user, and enable unobtrusive real-time monitoring by ensuring a more comfortable experience. One of the methods for extracting PR from video recordings is Eulerian Video Magnification (EVM), an algorithm that enhances subtle facial skin color variations to enable pulse visualization and analysis [13,14]. In the context of pulse detection, EVM enhances minor color fluctuations in facial skin that are correlated with blood circulation. The method involves spatial decomposition of each frame and temporal filtering to isolate and amplify specific frequency bands corresponding to physiological signals. By magnifying these subtle changes, EVM enables non-contact visualization and analysis of the pulse without requiring explicit tracking of facial features. A related approach is phase-based video motion processing [15], which allows for the manipulation of small movements relevant to physiological pulsations.

However, previous studies indicated that EVM may have limited accuracy in low-light conditions or when large head movements are present [1,14]. There are conflicting opinions regarding the feasibility of pulse extraction through video analysis. Various video-based methods have been developed for remote pulse measurement, including those based on head motion analysis [16,17], infrared imaging [2], and skin color changes recorded using RGB sensors [1,10,12]. Kwon et al. [10] conducted a study using a mobile phone camera for detecting facial color changes in 10 participants, demonstrating that the deviation from the reference electrocardiography (ECG) measurement was 2%. However, their study was conducted on a limited sample of participants and under controlled experimental conditions. In this context, Renner et al. [1] conducted a study with a larger number of participants in simulated driving conditions, concluding that pulse measurement using a low-cost webcam is unreliable in such conditions and that an alternative method for PR extraction is required. Here, we aim to answer whether it is possible to use video-based PR assessment during operation of a simulated vehicle in a healthy sample.

This paper introduces an rPPG method for extracting PR from video recordings captured during driving simulations using an affordable RGB web camera. The analysis is performed on a unique dataset comprising 79 recordings of 65 healthy individuals engaged in a driving simulation, spanning various age groups and a broad range of PR values, ensuring diverse data for evaluating the method. We would also emphasize that data used in this study were previously recorded for another purpose, and that camera was used solely with the purpose of supervision of measurement conditions. Therefore, our study is retrospective i.e., *a posteriori* in nature [18]. Thus, presented PR extraction constitutes a more challenging task in relation to PR assessment in a controlled environment (e.g., reduced subject's motion and dedicated camera focus).

The proposed method for extracting pulse from video recordings is based on detecting delicate changes in facial skin color captured using an RGB camera, which occur due to the blood volume changes in the microvascular tissue [6] that modulate the absorption and reflection of light. As a reference value, we use measurements obtained from the Empatica E4 sensor, which enables pulse tracking via a PPG sensor. By comparing the results of the proposed method with data from the Empatica E4 sensor, as a reference, we evaluate method effectiveness. Additionally, to enhance the accuracy of the extracted signal, we apply EVM. Then, we analyze whether the application of the EVM method improves the results compared to video analysis without EVM.

The measurements are conducted on participants from different age groups, allowing us to analyze potential differences in the extracted HR between younger and older groups of drivers. This study primarily investigates the agreement between PR derived from facial video and that obtained from the Empatica E4 device. Variations in experimental conditions, such as age group differences, are used as a framework to explore method reliability across different scenarios.

**Research Questions**

Specifically, we aim to address the following research questions:

1) Is it possible to estimate PR successfully in a driving simulator environment by application of a video-based method?
2) Would video-based PR assessment detect changes in PR caused by changes that exist between different age groups of participants?
3) Does the application of the EVM method contribute to the improvement of the PR assessment compared to the analysis without the application of the EVM?

We explore the feasibility of contactless monitoring of physiological parameters, such as PR, in specific driving simulation conditions. The focus is on identifying differences in pulse between older and younger groups of drivers. We conduct the analysis of video signals before and after applying the EVM method (B.EVM and A.EVM, respectively), with the goal of determining whether EVM contributes to the improvement of the pulse extraction accuracy or if its application is redundant, considering that it represents an additional and complex step in the analysis.

## 2. Materials and Methods

Figure 1 briefly summarizes the proposed algorithm for PR assessment. The complete code is implemented in Python version 3.11.0 (Python Software Foundation, Wilmington, DE, USA) within the Integrated Development Environment (IDE) Spyder [19], as well as in the MATLAB program version 2023a (The MathWorks, Natick, MA, USA). The following Python libraries are used for the realization of this research: NumPy [20], OpenCV [21], SciPy [22], Scikit-Learn [23], and math [24]. After loading the video recording, the detection of the subject's face is performed in each frame of the video recording. Following this step, the eyes are detected in the previously extracted facial region and only the region of interest (ROI) is kept, i.e., the subject's face without eyes. Once the ROIs are extracted for each frame, we create new videos from loaded frames, containing only ROI. The next step is the creation of 30 s long video sequences from the previously obtained video containing the ROI. Further, for each obtained video, in the first case, we apply EVM and extract the Signal of Change in Light Intensity (SCLI) that corresponds to the blood flow, while in the second case, we proceed with processing without applying EVM to extract SCLI. The final step involves the detection of peaks that correspond to the heart beats by applying a modified Pan–Tompkins method.

### 2.1. Dataset

Cardiovascular physiological signals are assessed by analyzing video recordings from two studies conducted in a compact, motion-based driving simulator. A total of 27 participants underwent repeated measurements in two different driving simulation environments, while one additional participant had only a single recording (without repeated measurements in the two different simulation environments). The first study was not designed to investigate physiological differences during driving simulations between age groups. Therefore, the overall number of recordings is $2 \times 27 + 1 = 55$ with 28 participants. The second study included 37 participants, divided into two age groups: 15 younger drivers (aged 30–45) and 22 older drivers (aged 60–75). Each participant in this study was recorded once during a single driving condition. Details related to the performed measurements and thorough technical specifications are given in [18]. The videos were obtained using a low-cost Logitech C920s

Pro HD Webcam which integrated auto focus, automatic lighting correction, and 78° field of view, positioned approximately 1.5 m away from the participants. The camera was positioned in front of the participant recorded videos in an RGB format and captured 30 frames per second (fps). The resolution of the videos belonging to the first study is 1280 × 720 pixels, while the resolution of those from the second study is 640 × 480 pixels. In addition, the dataset contains data from Empatica E4 (Empatica Inc., Boston, MA, USA) sensor obtained through simultaneous recording during the trials. Empatica E4 is a wearable device designed to monitor and collect data in real time [25,26]. This wearable sensor was placed on the subject's wrist (on their non-dominant hand to reduce motion artifacts) to measure, among other data, blood volume pulse (BVP) that we use to evaluate the proposed video-based method [25]. The Empatica E4 sensor, while widely used for physiological signal acquisition, can exhibit limitations in dynamic environments due to motion artifacts, loose contact, and sensitivity to ambient conditions such as temperature and humidity [25–27]. These factors can introduce noise and inaccuracies in the BVP and IBI signals, potentially affecting the reliability of measurements during activities involving significant subject movement [27]. BVP is an essential signal obtained from the Empatica E4 PPG sensor. The corresponding IBI and PR values are also generated by the device itself and were not computed in this study by ourselves.

Due to poor sensor–skin contact and subject movements, some recordings contain missing data. Unfortunately, we had to exclude videos for which reference signals from the Empatica E4 sensor (BVP, PR, and IBI) were not available. Overall, 17 videos are excluded from further analysis with 10 videos from the first study. For older and younger groups of drivers from the second study, simultaneous information is missing in six and one video, respectively. The sampling frequency of the BVP signal was 64 Hz

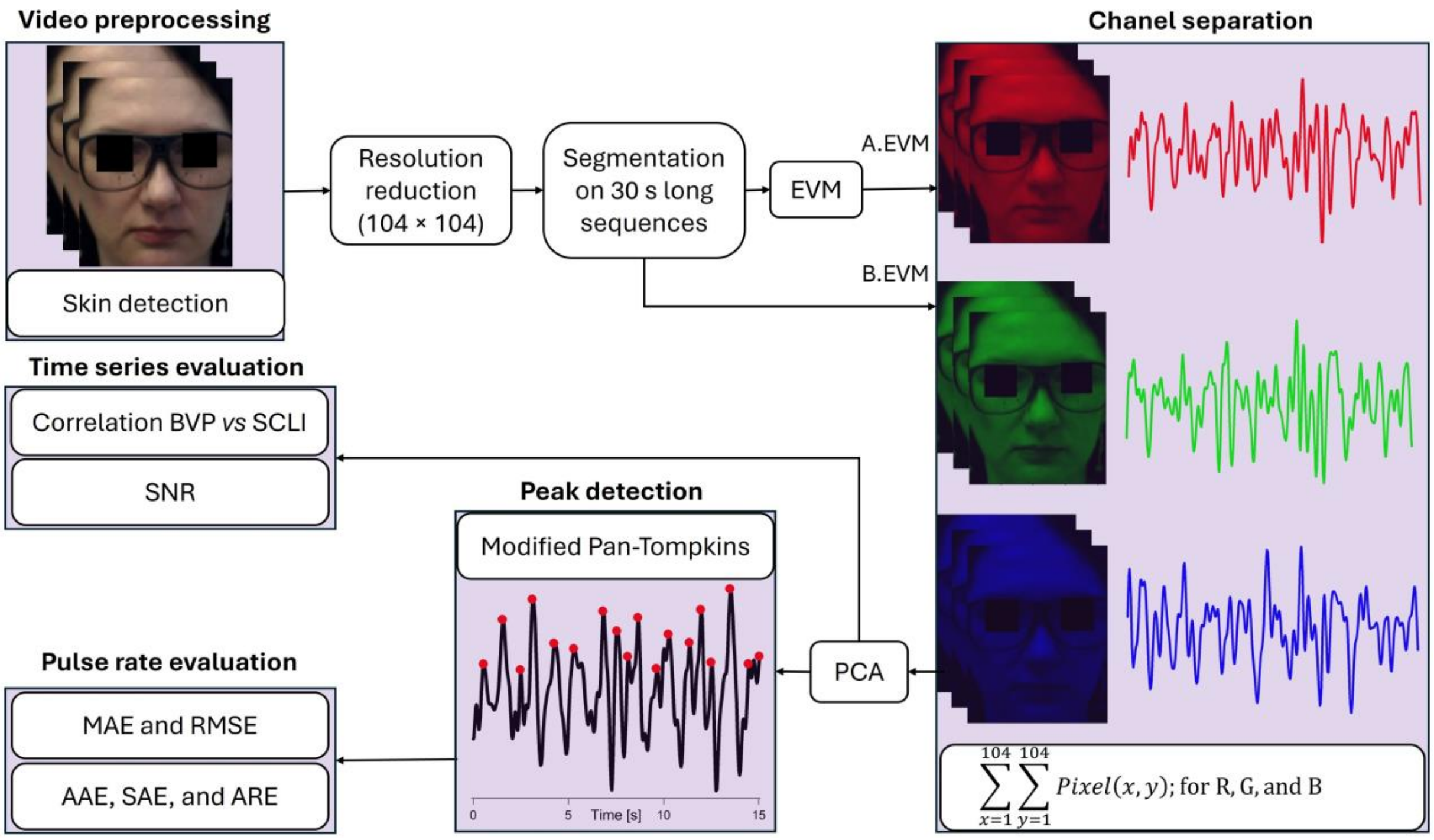

**Figure 1.** Block diagram of proposed algorithm for video-based pulse rate assessment. In the first case, SCLI is extracted B.EVM. In the second case, SCLI is extracted A.EVM. In the lower right corner, x and y refer to locations along the vertical and horizontal axes, respectively. Pixel refers to pixel value of red, green, and blue (R, G, and B) components at the given location. Abbreviations are: SCLI—Signal of Change in Light Intensity, B.EVM—Before applying Eulerian Video Magnification, A.EVM—After applying Eulerian Video Magnification, MAE—Mean Absolute Error, RMSE—Root Mean Squared Error, AAE—Average Absolute Error, SAE—Standard Deviation of Absolute Error, ARE—Average Relative Error, SNR—Signal to Noise Ratio.

### 2.2. Face Detection

After loading video recordings, a face is detected in each frame throughout the video (Figure 1). The face detection is performed using the You Only Look Once version 8 (YOLO v8) deep learning model [28]. First, the algorithm saves the input image. Further, the YOLO v8 model trained for face detection [28] is loaded and applied to the saved image. If no faces are detected (sometimes researchers entered in the frame, so two faces could be detected), the function returns the original image and the initial bounding box coordinates, which are manually set for the first frame. If faces are detected, then the face with the largest bounding box area is extracted. The final step involves the deletion of pixel values, i.e., assigning a value of 0 to each of the R, G, and B channels for all pixels that do not belong to the boundaries of the extracted rectangle around the face. Briefly, the algorithm returns the image and the updated coordinates of the largest detected face with the identified bounding box.

In addition to employing the YOLO v8 model, face detection is also explored using the Viola–Jones algorithm [29,30]. The implementation is utilized with the Haar cascade classifier [29,30]. However, under the specific recording conditions, such as when multiple individuals are present in the frame, especially with inconsistent lighting conditions and followed by expressed subject's movement, the Viola–Jones algorithm tended to misidentify faces. Examples include detecting the faces of researchers or mistakenly recognizing objects like light switches as faces due to their rough visual similarity to human features. Given these challenges, we opt to focus on face detection exclusively using the YOLO v8 model, which demonstrated a much lower incidence of fault during our initial exploration of algorithms for face detection.

### 2.3. Facial Skin Detection

Eye detection is performed as an initial step in the facial skin extraction procedure. Our assumption is that by detecting eyes and excluding them from the face images, we may extract facial skin, which is essential for video-based PR detection [31]. The algorithm for eyes detection accepts an RGB color input image (with a previously detected face region), converts it to grayscale, loads a predefined Haar cascade classifier model for eye detection [30], and applies the model to the grayscale image to detect eyes. Grayscale conversion is a typical practice in many image processing algorithms as it simplifies processing and reduces the amount of data [32]. Since a pretrained YOLO model [28] was available only for face detection and not for eyes, we used the Haar cascade classifier for eye detection, as it is publicly available and specifically trained for that purpose [33].

Haar cascade classifiers with a scale factor of 1.01 (as recommended in [34]) allow for fine scaling of the window of interest, ensuring a more precise detection of characteristic features, like eyes in our case. Further, setting the number of neighboring rectangles to three [34] reduces false positives by requiring multiple detections in neigh-boring areas before confirming the presence of the detected feature. We use these parameters, due to their proven efficacy: they balance accuracy and computational efficiency, providing reliable detection in challenging low resolution conditions [34]. The classifier returns a list of rectangles, where each rectangle represents an ROI containing a detected eye, defined by four values: the horizontal and vertical coordinates of the top-left corner, as well as the width and the height of the rectangle. In each frame, we expect exactly two ROIs to be detected.

In the case where only one eye is detected, the position of the other eye is assumed relative to the longitudinal axis of symmetry set at the center of the facial ROI. If the system cannot detect at least one eye after extracting the face, the eye regions detected in the previous frame are retained. The rectangles representing the detected eye regions are excluded from the facial region by assigning a value of 0 (turning black) to the corresponding pixels. After extracting the eyes, all black pixels not belonging to the facial region are excluded, meaning the image is cropped. Finally, the image is resized to dimensions of $104 \times 104$ pixels to obtain frames of identical dimensions, which is required for further application of EVM.

In the videos where the facial features of the participants are extracted, to effectively analyze physiological signals, we divide the entire video, which lasts approximately 20 min, into sequences of 30 s duration [35,36]. Each new sequence begins 10 s after the previous one, meaning there is an overlap of 20 s between adjacent sequences. In the following analysis, adjacent sequences of a 20 s span ensure a smooth transition between analysis windows as previously proposed in [35,36]. This overlapping method ensures that we capture continuous signal variations and reduce the chance of missing critical information. According to [37], different interval durations can be successfully used for PR estimation, such as long-term (including 24 h recording or longer), short-term (approximately 5 min), or even ultra-short-term (less than 5 min), while several publications [38–41] state that 10 s is sufficient for a successful PR estimation. In addition, intervals lasting 10 s were used in the work [12], in which the non-contact evaluation of the pulse is determined based on the change in intensity of the RGB components. The similarity in methodological approaches and interval durations reported in related studies supports the validity of our selected windowing strategy. While 10 s windows are commonly used [42], we chose a different interval, as it provides more robust pulse extraction in the presence of substantial lighting changes and participant motion.

### 2.4. Application of Eulerian Video Magnification

EVM is applied on overlapping video sequences, containing only frames with selected ROIs (on faces without eyes, i.e., facial skin). By emphasizing tiny changes in facial color that may be indications of physiological signals, the EVM method makes variations in video recordings that would otherwise be invisible [13,15]. In this paper, we want to examine the effectiveness of EVM on PR extraction by evaluating PR obtained from B.EVM and A.EVM. Our main motivation lies in the fact that while EVM proved its usability [43], it can be time consumable, and its application may not be required for pulse assessment [42].

The analysis involving the EVM approach is performed in MATLAB by application of the available code examples from Wu et al. [13] and Wadhwa et al. [15]. After that, Gaussian and Laplacian pyramids are formed with two levels and with a depth of three because it does not make sense to further reduce the size of an image that is initially 104 × 104 pixels, which after applying the Gaussian and Laplacian pyramids constitutes of only 13 × 13 pixels. These pyramids reduce noise influence and allow for more efficient signal processing, as lower-resolution images can be processed more quickly [30].

Then, a temporal filter is applied to the time series comprising intensity values of each pixel across time to isolate the frequency components of interest. This filter is created by transforming the time signal into the frequency domain using the fast Fourier transform (FFT), where all frequency components below 0.4 Hz and above 3 Hz are set to 0 [13,15] (which corresponds to a PR range of 24 beats per minute (bpm) to 180 bpm) as recommended in [13]. The frequency range of 0.4 Hz to 3 Hz is selected to cover PR from 24 bpm to 180 bpm, which encompasses normal and elevated PR in healthy individuals [44]. This range is slightly narrower than the one used in similar studies (e.g., Wu et al. [13] (0.4 Hz–4 Hz); Wang et al. [45] (0.4 Hz–4 Hz)), but it is chosen to focus on physiologically relevant signals, avoiding potential noise influence. After that, the signal is transformed back into the time domain, and the filtered signals are multiplied by a magnification factor of 20 to increase the amplitude of pixel values in the frames. While Wu et al. [13] suggest using a higher magnification factor value of 120 to emphasize subtle skin color changes, we use an empirically selected amplification factor of 20 that can still amplify subtle skin color changes caused by blood flow through blood vessels, without going into pixel value saturation, which is the case when we initially applied a factor of 120 to our videos. Finally, the reconstruction of the video is achieved by adding the amplified components to the original video, allowing the visualization of subtle changes such as pulse-induced skin color variations. This reconstructed video is then used for further analysis and signal extraction (e.g., average intensity calculation over the region of interest), as illustrated in Figure 1.

Due to the recording protocol that envisioned slight subject movements, we decided not to apply the phase motion processing method, although phase motion processing can effectively freeze small movements [15]. Despite the fact that phase motion processing is applied to eliminate motion artifacts combined with EVM [15], in our case, relatively large head movements lead to considerable blurring and the appearance of noise in the video frames when phase motion processing is applied. We choose to use the EVM method without adding a phase-based motion processing step because these large head movements typically cause problems with reliable PR analysis and monitoring, which can hinder accurate analysis and PR measurement [15].

We conduct a time complexity analysis of the EVM method to quantify its computational demands [46,47] across varying video durations and execution settings. The code execution and performance measurements were conducted on a laptop equipped with an 11th Generation Intel(R) Core

i9-11900H processor operating at 2.50 GHz, with 16 GB of random access memory. The operating system used was Windows 11 Home, 64-bit.

Specifically, we select video recordings with durations of 30 s, 1 min, 5 min, 10 min, 15 min, and 20 min to represent a range of realistic use cases in practical applications. For each of these durations, the EVM processing pipeline was executed 10 times to obtain stable and reliable measurements of both execution time, mitigating the impact of potential fluctuations caused by background processes and system variability.

Additionally, we systematically examine all scenarios under two execution settings: (1) Single-core execution: The EVM algorithm is executed using only one processing core to simulate scenarios on computationally limited devices or where parallelization is not feasible. (2) Four-core execution: The same EVM processing was executed using four processing cores to leverage parallelization capabilities, representing typical high-performance computing environments or real-time analysis pipelines with available multi-core resources.

### 2.5. Extraction of Light Changes and Peak Detection

After extraction of the facial skin region, the pixels corresponding to the eyes are excluded by setting their values to zero. Then, for each frame in the overlapping video sequence, the mean pixel value is calculated separately for each of the three color channels (R, G, and B). This results in a time-varying signal for each color channel, with as many samples as there are frames in the video. These signals are further filtered (if EVM is not applied) to suppress components unrelated to cardiac activity. A third-order Butterworth band-pass filter was applied, with cutoff frequencies set to 0.4 Hz (corresponding to a pulse of 24 bpm) and 3 Hz (corresponding to 180 bpm). The upper and lower filtering limits are chosen to correspond to the physiological PR limits during moderate exercise [44]. In both cases, filtering is applied in both directions to ensure zero-phase filtering.

According to the results from previous studies [10,11], after extracting the R, G, and B components, the application of independent component analysis (ICA) or principal component analysis (PCA) is advised to more effectively segregate the desired information from the noise. As proposed in [9], PCA is a less computationally demanding method than ICA, and the accuracy of the extracted pulse using these two methods is very similar. It has been shown that the first principal component contains most of the information, while the other components have increasing amounts of noise [9,48]. Thus, we use only the first principal component for further analysis. We assume that the color changes would be the most dominant and the most variable component in comparison to, for example, varying external lighting conditions during the recording session. To mitigate the impact of pronounced peaks due to unwanted subject movements, values deviating for more than three interquartile ranges from the third quartile are considered outliers and subsequently adjusted [49]. The resulting signal is termed SCLI.

Extraction of the SCLI is performed in two ways (Figure 1). The result of applying EVM to the video sequence is also a video sequence with more pronounced color changes in the face corresponding to the blood flow. The method for extracting SCLI is the same in both cases (B.EVM and A.EVM); the only difference is the video sequence that is used. In the B.EVM approach, the sequence is processed after facial skin extraction, whereas in the A.EVM approach, the EVM method is first applied to the sequence containing the extracted facial skin.

To detect the peaks of the SCLI that correspond to PR pulsations, the modified Pan–Tompkins (PT) algorithm is used [41,50]. Since the signal has already been filtered, this step in the modified PT method is simply skipped so as not to filter it multiple times. The first derivative is applied on the SCLI to enhance peaks. Then, SCLI is filtered by the application of the moving average filter. The search for the moving average window width is conducted with the widths that varied from 33 ms to 1 s with the step of 33 ms (corresponds to averaging from one to 30 samples because the sampling frequency is 30 Hz) [51]. Suitable values of the window width are determined for each overlapping video sequence to minimize the mean absolute error (MAE) in comparison with the reference average PR values from the Empatica E4 sensor in the corresponding sequence. All suitable values of the moving average window widths are then averaged to obtain a unique parameter applicable to all videos. The rest of the modified PT algorithm uses a simple method based on the single threshold application for peak detection. The *findpeaks* function with a minimum peak distance parameter of 0.33 s is applied since we assume that healthy subjects will not have a higher PR than 180 bpm in a sitting position during driving simulation [52]. In the signal analysis, the prominence parameter is used instead of a fixed threshold for peak detection to ensure method robustness in the presence of both prominent dominant peaks and smaller peaks that may represent noise. Specifically, the analyzed signals contain well-defined dominant peaks as well as smaller peaks, the amplitude of which, although distinguishable, varies significantly relative to the dominant peaks. Setting a static threshold based on signal height would be unreliable, as the threshold value could either eliminate significant peaks or retain excessive noise components. In the peak detection process, we use a prominence threshold of 0.15, which refers to the minimum required height difference between a peak and its surrounding baseline. This value is expressed in absolute units of the signal amplitude, making this approach more suitable for analyzing signals with variable dominant peaks [53,54]. The specific value of 0.15 is empirically selected.

### 2.6. Evaluation Metrics

The linear and monotonic correlations between the corresponding BVP (from the Empatica E4 sensor) and SCLI (obtained from both B.EVM and A.EVM) are evaluated in this work using the Pearson and Spearman correlation coefficients, respectively. Spearman correlation assesses monotonic links, making it less vulnerable to nonlinear associations or outliers than Pearson correlation, which analyzes linear dependency [55]. Ideally, we would expect a linear relationship between BVP and SCLI; however, due to the measurement locations (wrist and face), different sensing modalities, divergent noises, and dissimilar processing techniques, this may not be guaranteed. To enable a valid comparison between the BVP signals recorded by the reference Empatica E4 device and those extracted from the corresponding video sequences, the sampling rates are adjusted by applying cubic interpolation for up sampling of signal extracted from video, so that the time series would contain the same number of data points.

To calculate the signal-to-noise ratio (SNR), two different types of signals are used: raw BVP from the reference Empatica E4 device and the signal obtained from the video recording after applying PCA. These signals were not filtered before. First, the reference BVP signal from the Empatica E4 sensor is filtered using the Butterworth zero-phase third-order digital band pass filter in the range from 0.4 Hz to 3 Hz to extract frequency components corresponding to PR. After filtering, the signal power is calculated as the square value of the filtered signal. Noise is defined as the difference between the filtered and raw BVP signals by simple subtraction in time domain, while the noise power is calculated as the mean square value of the noise. For SNR estimation for the signal obtained from the video recording, the filtered signals from the R, G, and B color channels are stacked into a matrix, and PCA is applied. Only the first principal component is retained, representing the most dominant variation shared across the three channels (R, G, and B). The first PCA component is stacked into a matrix for further comparison with R, G, and B components [56]. The noise is calculated as the difference between the filtered R, G, and B components and first component of PCA analysis. The signal and the noise power are calculated as the mean square value of the estimated noise and measured signal, respectively. The calculated SNR values for both cases (B.EVM and A.EVM) are stored for further analysis.

We use the obtained PR and IBI signals from Empatica E4 sensor as the ground truth in the process of evaluation of video-based PR assessment. To understand the difference between the data from the reference Empatica E4 sensor and the suggested approach, given in bpm, the extracted PR values are used. PR values are computed in 30 s segments by detecting peaks within each segment and calculating the average IBI *(IBIaverage*). PR is then derived as the inverse of this average interval using the equation: $PR = 60/IBIaverage$. To compare PRs from both reference device and video, the MAEs and root mean square errors (RMSEs) are computed for each extracted segment. The MAE represents the average absolute difference between the corresponding PR values from Empatica E4 sensor and extracted PR from the SCLI. A smaller MAE indicates a lower overall error, reflecting a higher level of agreement between the two devices. On the other hand, RMSE represents the square root of the averaged squared differences between two values, placing greater emphasis on larger errors compared to MAE. In addition to the standard metrics of MAE and RMSE, which are commonly used to evaluate the success of pulse extraction from camera videos [1,57,58], we also employ absolute error (AE), average absolute error (AAE), standard deviation of the absolute error (SAE), and average relative error (ARE), recommended in [59] and compute them for each segment by applying Equations (1)–(4). These additional metrics include deviations in absolute values, their variability, as well as the relative error with respect to the reference values, thus providing a more comprehensive insight into the accuracy of the results obtained. *PRreference* and *PRextracted* stand for reference PR from Empatica

E4 device and extracted PR from the video method, while $N$ is the overall number of video sequences and i indicate single instances, i.e., videos.

The lack of accessibility of databases captured during the driving simulation is an issue, so we expand evaluation on immediately publicly available databases discussed in [60]. Three publicly available datasets [61,62] were initially considered for this study. However, the dataset by Benezeth et al. [62] is excluded because the videos were obtained with a near-infrared camera. From the remaining two datasets [61], we pseudorandomly select five video recordings with the goal of including at least 5 subjects and at least 25 sequences of 30 s per dataset. Both videos datasets published in [61] were captured under controlled conditions, which generally resulted in limited subject movements. However, in one case, participants were squinting during the recordings, which posed a challenge for our eye detector [63].

$$AE(i) = |PR_{reference}(i) - PR_{extracted}(i)| \quad (1)$$

$$AAE = \frac{1}{N} * \sum_{i=1}^{N} AE(i)\ AE \quad (2)$$

$$SAE = \sqrt{\frac{1}{N} * \sum_{i=1}^{N} (AE(i) - AAE)^2} \quad (3)$$

$$ARE = \frac{1}{N} * \sum_{i=1}^{N} \sum_{i=1}^{N} \frac{AE(i)}{PR_{reference}(i)} \quad (4)$$

### 2.7. Additional Processing of Extracted Pulse Rates

A noticeable increase in the deviations (assessed by all evaluation parameters) presented in Figure 2 is observed with the increase in reference PR extracted from the Empatica E4 device. This observation aligns with the results presented by Bent et al. [64], showing that MAE increases after physical activity. Stuyck et al. [65] advise researchers to apply correction factors, as they observed that Empatica E4 overestimates PR compared to ECG. In Figure 2, it can be observed that this increase when PR extracted from video recordings and compared to the reference PR obtained from Empatica E4 follows an approximately linear trend. Therefore, we decide to design a linear fit (showcased on the left-hand panel graphs in Figure 2). Based on this, we correct the extracted faults by subtracting the linear fit values from the calculated differences between the reference and extracted PR (linear fit correction).

For the sake of scientific rigor and with the aim to avoid any possible biases that may be introduced by video-based PR assessment, we use data shared with us by Medarević et al. [66]. Thus, we assess linear fit related to the PR increase when Empatica E4 measurements are compared against PR measurements from the Faros 360 [67]. A similar trend is observed between Empatica E4 and Faros 360, although the correction of such a linear trend introduced less changes in our dataset (graphs on right-hand panels in Figure 2). In our case, a simple linear fit correction is applied, with parameters a and b in the basic linear equation $y = a * x + b$, which are 0.94 and −69.41 for B.EVM as well as 0.96 and −74.01 for A.EVM, respectively, effectively reducing observed deviations. A similar linear trend was recorded between the data from Empatica E4 and Faros 360 devices presented in [66], where the parameters of the linear fit were 0.32 and −30.42. The left-hand panel in Figure 2 presents the linear fit designed onto our data. The differences between the extracted PR from the video recording and the reference PR from Empatica E4, in both B.EVM and A.EVM cases, are shown as dark circles. Lighter circles represent the corrected differences between reference and extracted PR after subtracting the linear fit from the data depicted by the black circles. Further analysis is performed on both corrected (linear fit correction designed for our data and linear fit correction based on data from [66]) and uncorrected values to examine the effectiveness of eliminating deviations between reference and extracted PR values.

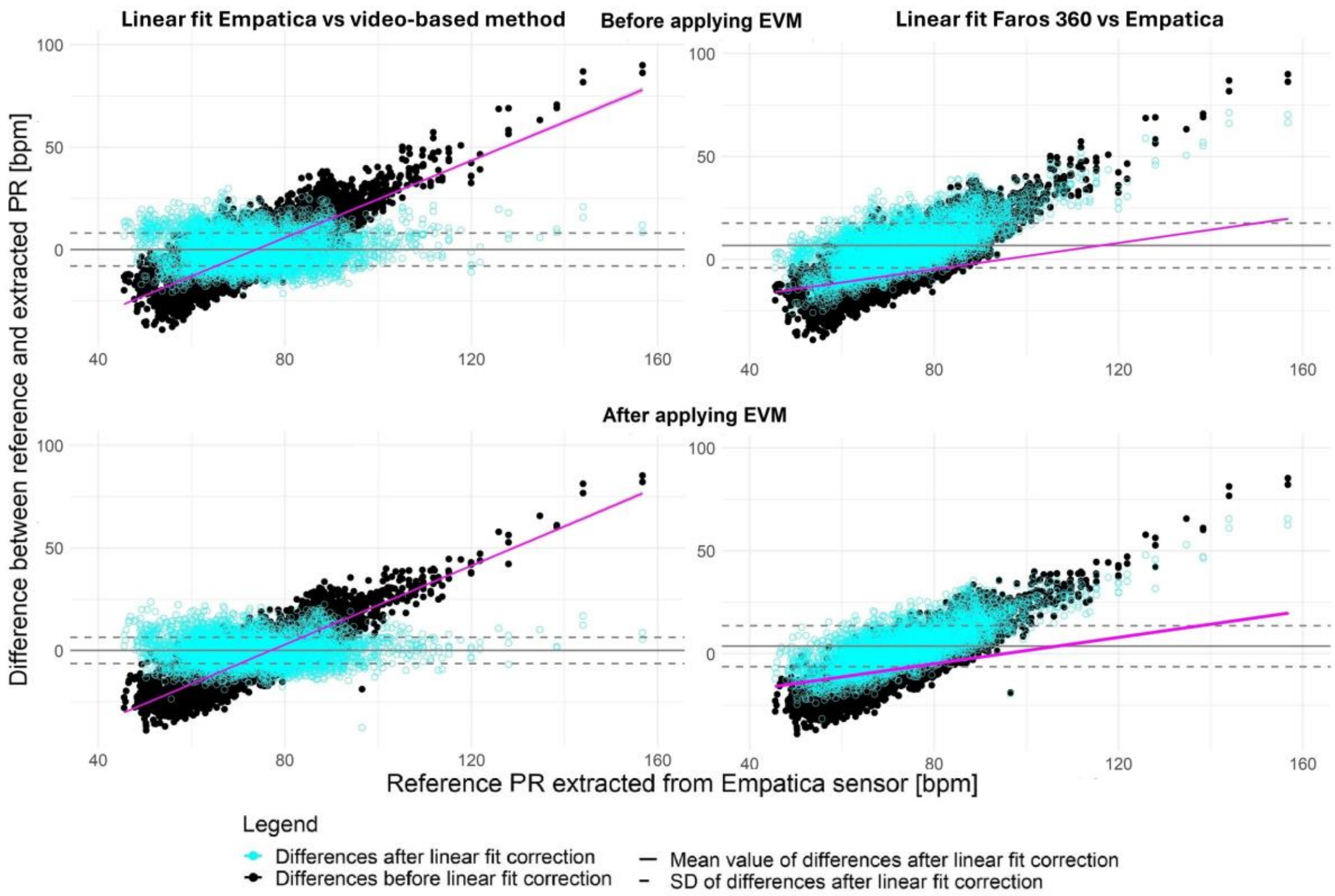

**Figure 2.** The increase in differences between reference and extracted PR follows a linear trend (magenta color). The graphs on the upper panel represent the B.EVM case, and the lower one represents the A.EVM case. Gray lines refer to the mean, and dashed lines to the standard deviation. Magenta lines refer to the corresponding linear fit B.EVM—Before applying Eulerian Video Magnification, A.EVM—After applying Eulerian Video Magnification.

### 2.8. Statistical Tests

In the conducted analysis, statistically significant differences in the averaged PR values obtained for two age groups of participants are examined. An analysis is performed between the older and younger groups of participants who took part in the second study.

The normality of the PR distribution in each group is tested using the Shapiro–Wilk test. If the Shapiro–Wilk test indicates that the data follow a normal distribution an unpaired t-test is used for the older and younger groups of participants in the second study because the respondents are mutually independent. If the data do not follow a normal distribution, the non-parametric Wilcoxon signed-rank test is applied. The significance threshold is set at 0.05.

Additionally, effect size is assessed to evaluate the practical significance of PR differences between groups. This is important, as statistically significant results can sometimes arise solely due to large sample sizes [68]. To quantify the effect size, Cohen's *d* is used if the data followed a normal distribution. Cohen's *d* values are interpreted as small (approximately 0.2), medium (approximately 0.5), or large (0.8 and above) [68]. If the data does not follow a normal distribution, Cliff's *delta* is performed. Cliff's *delta* values range from −1 to 1, where extreme values indicate complete separation between groups, while values near zero suggest very small or negligible differences [68].

## 3. Results

In Figure 3, we present time-series obtained together with spectral density, which is estimated using the periodogram with Bartlett smoothing, from a single subject participating in the first study. These include the reference BVP signal, as well as the SCLI B.EVM and SCLI A.EVM signals extracted from the video recording.

When SCLI is calculated without the use of EVM, the average Pearson and Spearman correlation coefficients across participants are 0.08. With the use of EVM, these values slightly increase to 0.09. The standard deviations are similar to the mean values. For the Pearson correlation coefficient, the standard deviations are 0.06 for B.EVM and 0.07 for A.EVM. In the case of the Spearman correlation coefficient, the standard deviations are slightly higher, amounting to 0.10 and 0.11, respectively, for the conditions without and with EVM. The Pearson cross-correlation coefficient between the SCLI B.EVM signal and the reference BVP signal on the sequence shown in Figure 3 is 0.12, while it decreases to 0.09 after applying the EVM method. Similarly, the Spearman cross-correlation coefficients are 0.16 for B.EVM and 0.13 for A.EVM. These deviations from the average values are likely attributed to the calculation being performed on the short sequence presented in Figure 3.

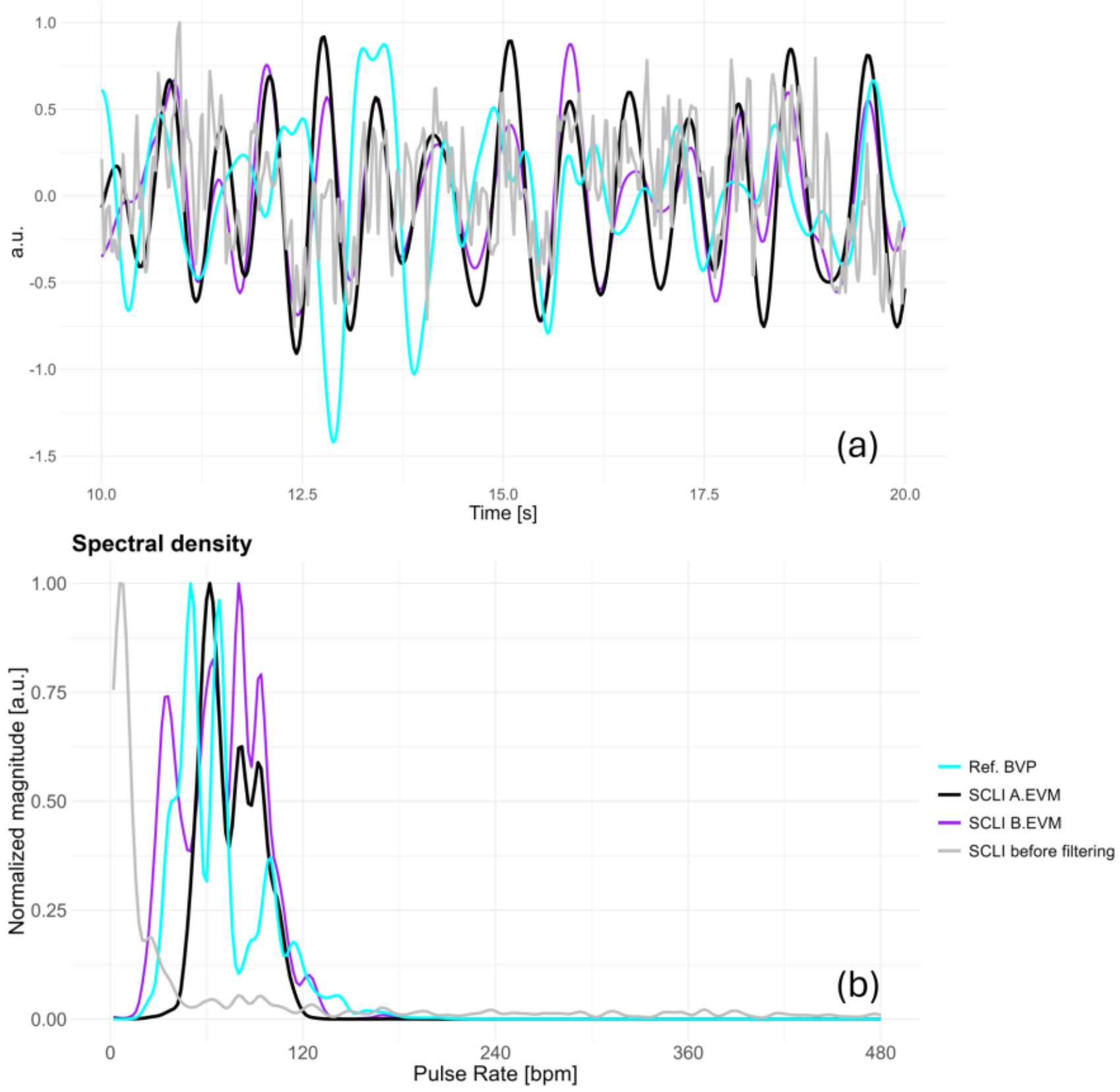

**Figure 3.** Three time-series that display the reference BVP signal and SCLI without filtering. SCLI B.EVM and A.EVM cases obtained on a sample participant from the first study are presented in panel (**a**). Corresponding spectral density is illustrated in panel (**b**). B.EVM—Before applying Eulerian Video Magnification, A.EVM—After applying Eulerian Video Magnification, BVP—Blood Volume Pulse, SCLI—Signal of Change in Light Intensity.

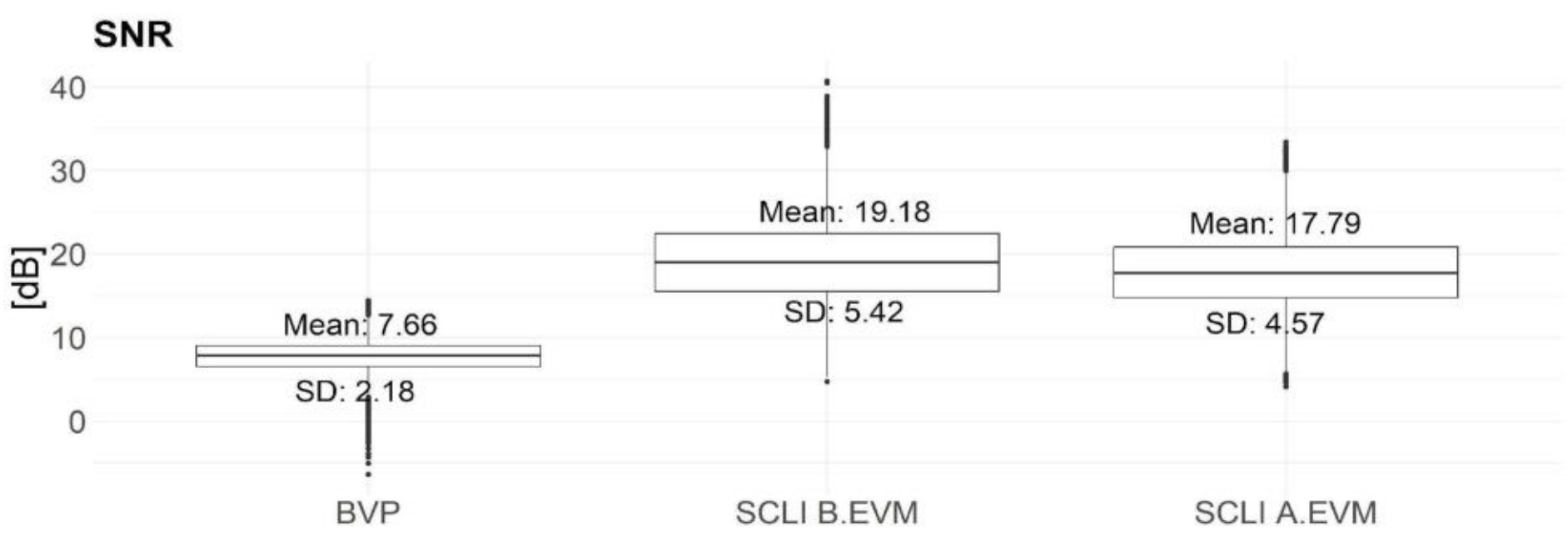

**Figure 4.** SNRs calculated across all video sequences for recording obtained from all subjects for BVP, SCLI B.EVM, and SCLI A.EVM. B.EVM—Before applying Eulerian Video Magnification, A.EVM—After applying Eulerian Video Magnification, SNR—Signal to Noise Ratio, BVP—Blood Volume Pulse, SCLI—Signal of Change in Light Intensity.

The average and standard deviations of the SNR values for corresponding BVP signals obtained from Empatica E4 sensors and SCLI (B.EVM and A.EVM), computed for all video sequences of all subjects, are shown in Figure 4. The results demonstrate that in all measurements, SCLI has a higher average SNR than BVP signal.

The width of the moving average window, which results in the lowest MAE values, is 400 ms (corresponding to 12 samples) for SCLI B.EVM and 433 ms (corresponding to 13 samples) for SCLI A.EVM. The reference PR values obtained from the Empatica E4 and extracted PR values from the SCLI (B.EVM and A.EVM) are compared for each video sequence for all participants. Before applying the linear fit correction, calculated MAE and RMSE values (B.EVM and A.EVM) are shown in the upper panel of Figure 5. The results show that for the B.EVM case before the linear fit correction, the average AAE for all sequences and participants is 10.46 bpm, while the average SAE is 8.83 bpm. Additionally, the average ARE is 0.14 (14%). For the A.EVM case, the average AAE is 10.55 bpm, the average SAE is 8.29 bpm, and the average ARE value is 0.15 (15%). MAE and RMSE after linear fit correction (designed onto our data) are presented in the middle panel of Figure 5. The results demonstrate that, after linear fit correction incorporated into our data, the average AAE is 6.61 bpm, while the average SAE is 4.54 bpm. Additionally, the average ARE is 0.09 (9%). The average AAE decreases to 5.09 bpm due to the application of the EVM method, the SAE decreases to 3.93 bpm, and the ARE decreases to 0.07 (7%). Bottom panel of Figure 5 presents MAE and RMSE when the linear fit shown in [66] is applied to our data.

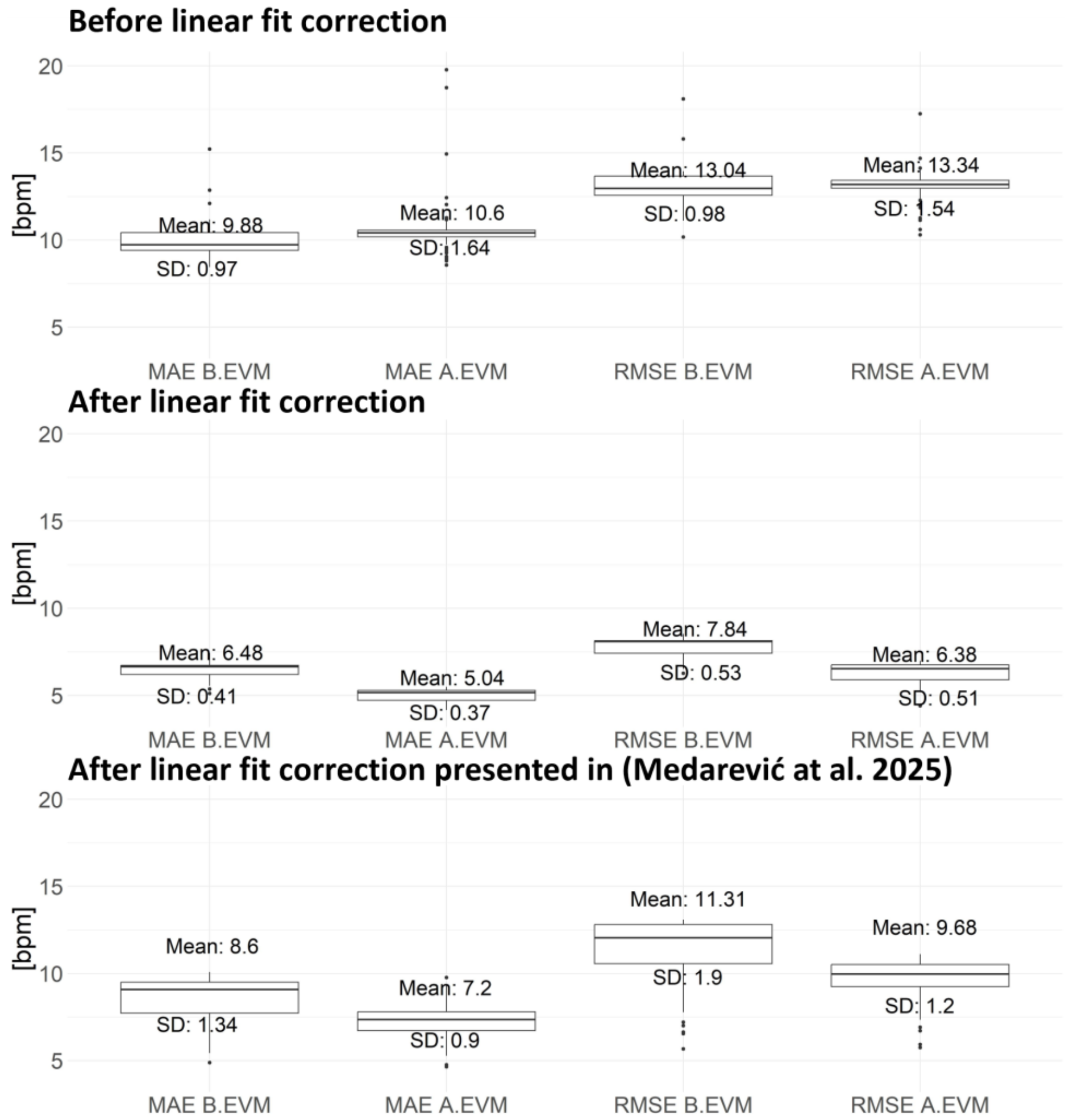

**Figure 5.** Comparison of PR values obtained from the Empatica E4 sensor and PR extracted from the SCLI B.EVM and the SCLI A.EVM before and after linear fit correction. The upper panel presents MAE and RMSE before linear fit correction, while lower panel displays MAE and RMSE after linear fit correction. MAE—Meana Absolute Error, RMSE—Root Mean Square Error, B.EVM—Before applying Eulerian Video Magnification, A.EVM—After applying Eulerian Video Magnification [66].

In Table 1, we compare the results of our approach on two publicly accessible datasets [61]. Additionally, we present the performance outcomes of approaches in literature that used ICA-based, POS, or CHROM-based techniques [11,45,57,61,69,70].

**Table 1.** Comparison of the proposed approach with other methods. The best result refers to the lowest average mean value among applied methods in our proposal. The lowest values from each column are bolded. MAE—Mean Absolute Error, RMSE—Root Mean Square Error, AAE—Average Absolute Error, SAE—Standard Deviation of Absolute Error, ARE—Average Relative Error, B.EVM—Before applying Eulerian Video Magnification, A.EVM—After applying Eulerian Video Magnification, ICA—Independent Component Analysis, PCA—Principal Component analysis, 3DCNN—Three Dimensions Convolutional Neural Network.

| Method | Simulation Environment | MAE [bpm] | RMSE [bpm] | AAE [bpm] | SAE [bpm] | ARE [%] |
|---|---|---|---|---|---|---|
| The best results presented in our paper | Driving simulator | 5.04 ± 0.37 | 6.38 ± 0.51 | 5.09 | 3.93 | 7.00 |
| Our approach applied to first dataset presented in [61] | Still with natural face expression | B.EVM: 9.52 ± 6.54<br>A.EVM: 9.90 ± 8.37 | B.EVM:10.53 ± 6.38<br>A.EVM: 10.26 ± 8.33 | / | / | / |
| Our approach applied to the second dataset presented in [61] | Still with natural face expression | B.EVM: 3.52 ± 0.84<br>A.EVM: 4.48 ± 2.09 | B.EVM: 4.33 ± 1.03<br>A.EVM: 4.84 ± 2.13 | | | |
| ICA is applied in [11] | Still | / | 4.63 | / | / | / |
| 3DCNN is applied in [36] | Still with natural face expression | **2.09** | 7.30 | / | / | / |
| ICA is applied in [45] | Still | / | 12.23 | / | / | / |
| CHROM is applied in [57] | Well controlled | 13.49 | 22.36 | / | / | |
| 3DCNN is applied in [57] | Well controlled | 5.96 | 7.88 | / | / | / |
| POS is applied in [61] | Still with natural face expression | / | 6.77 | / | / | / |
| CHROME is applied in [61] | Still with natural face expression | / | **2.39** | / | / | / |
| [71] | Well controlled and toothbrushing | 4.99 | / | / | / | / |
| ICA applied in [69] | Still with natural face expression | 8.83 | 12.24 | / | / | / |
| POS applied in [69] | Still with natural face expression | 5.76 | 9.67 | / | / | / |
| Green channel is used in [72] | Still with natural face expression | / | 8.35 | / | / | / |
| [73] | Still with natural face expression | / | / | **2.79** | 5.17 | **3.89** |
| [74] | Slightly rotation of the head | / | / | 9.89 | 4.23 | / |
| PCA is applied in [70] | Still with natural face expression | 5.42 | 6.13 | / | 4.28 | / |
| ICA is applied in [70] | Still with natural face expression | 5.66 | 6.48 | / | **3.59** | / |

When observing PR values obtained from two groups of subjects (older and younger groups of drivers in the second study), a statistically significant difference is recorded based on the reference data. The data did not follow a normal distribution, according to all of our tests. After extracting the PR from the video recordings, a statistically significant difference between the two groups (older and younger groups of drivers) is observed in both cases (both with B.EVM and A.EVM). This difference remains statistically significant even after applying a linear fit correction based on data from the Faros 360 and Empatica E4 sensor. Table 2 summarizes the *p* values obtained from the statistical tests, confirming significant differences between the two age groups under driving simulation conditions.

**Table 2.** Statistical tests indicate no normal distribution. Abbreviation *p*.f. refers to the *p* value after applying the linear fit correction and *p*.f.f. refers to the *p* value after correction using linear fit that is calculated from the data in [66]. B.EVM—Before applying Eulerian Video Magnification, A.EVM—After applying Eulerian Video Magnification. SCLI—Signal of Change in Light Intensity.

| **Older vs. Younger Groups of Drivers** | **Tests for Examining Statistically Significant Differences Wilcoxon Rank-Sum Test (*p* Value)** | | | **Effect Size Cliff's *Delta* Test** | | |
|---|---|---|---|---|---|---|
| | *p* | *p*.f. | *p*.f.f. | *p* | *p*.f. | *p*.f.f. |
| Reference Empatica E4 | <0.001 | / | / | 0.37 | / | / |
| SCLI B.EVM | 0.04 | <0.001 | <0.001 | 0.05 | 0.29 | 0.38 |
| SCLI A.EVM | 0.01 | <0.001 | <0.001 | 0.06 | 0.34 | 0.37 |

Figure 6 shows the PR values extracted for two groups of participants (older and younger groups of drivers), after linear fit correction designed onto our data. PR ranges by age categories (older and younger) from the literature are marked in gray [75]. According to the values obtained from the reference BVP signal from Empatica E4 sensor, a difference in the average PR can be observed (for the older group, the average extracted pulse value is 73.12 bpm, while for the younger group, it is 80.99 bpm). Statistical tests presented in Table 2 indicate a statistically significant difference between these two groups of participants based on the reference BVP signals and also based on SCLI (B.EVM and A.EVM), with and without linear fit corrections.

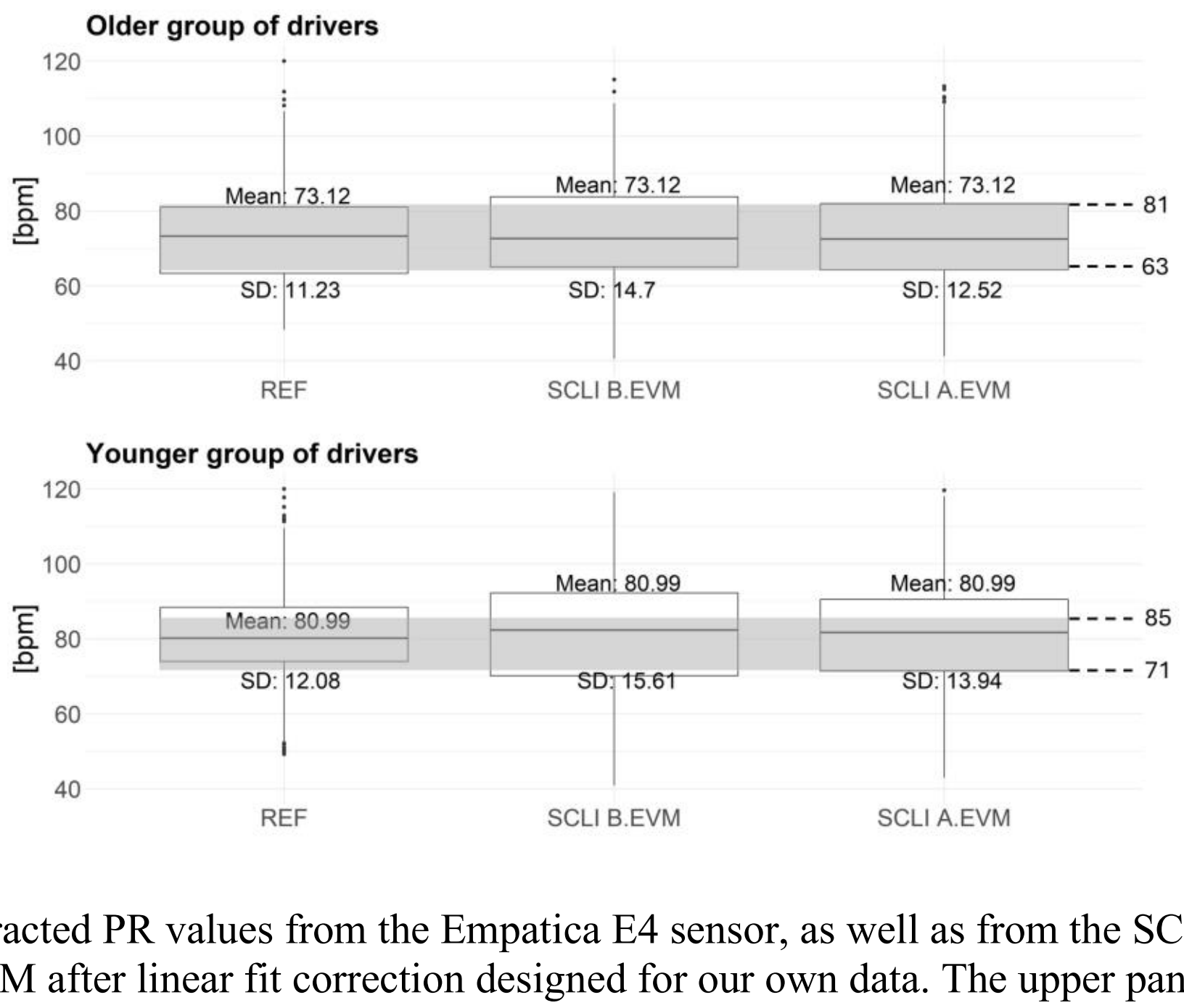

**Figure 6.** Extracted PR values from the Empatica E4 sensor, as well as from the SCLI B.EVM and SCLI A.EVM after linear fit correction designed for our own data. The upper panel shows the extracted values for the younger group of drivers belonging to the second study and the bottom panel shows the results of the older group in the second study. B.EVM—Before applying Eulerian Video Magnification, A.EVM—After applying Eulerian Video Magnification, SCLI—Signal of Change in Light Intensity.

The results of the analysis of time complexity when applying the EVM method are shown in Table 3. This clearly illustrates how the duration of the video and the execution mode (single-core or four-cores) affect the total processing time during the process.

**Table 3**. Execution time of the additional processing step in our method using single-core and four-cores parallel implementations.

| **Video Duration [min]** | **Single-Core** | **Four-Core** |
|---|---|---|
| | **Execution Time [s]** | **Execution Time [s]** |
| 0.5 | 23.26 ± 0.86 | 16.61 ± 1.62 |
| 1 | 53.85 ± 1.06 | 33.87 ± 0.85 |
| 5 | 99.78 ± 0.98 | 60.82 ± 1.36 |
| 10 | 394.69 ± 0.85 | 98.11 ± 1.29 |
| 15 | 720.12 ± 112.59 | 134.09 ± 0.98 |
| 20 | 998.84 ±11.68 | 176.76 ± 34.29 |

## 4. Discussion

The correlation coefficients between BVP signals obtained from Empatica E4 sensor and SCLI signals extracted from videos (for B.EVM and A.EVM cases) are reasonably low. The average values of Pearson and Spearman correlation coefficients are 0.08 when SCLI is obtained without the application of EVM and 0.09 when EVM is applied. In addition, the values of the standard deviations are almost identical to the mean values, while the standard deviations in the case when the Spearman correlation coefficient is calculated are slightly higher for both methods (B.EVM and A.EVM), which could indicate lower consistency and reliability as a consequence of the dynamic movements that are inevitable in a driving simulator. Relatively low cross-correlation coefficients may partially result from motion artifacts specific to the Empatica E4 device [57], such as wrist movement, which do not affect the video signal. It is important to note that head motion, which may influence video-based measurements, does not impact the PPG signal recorded by the wrist-worn Empatica E4 sensor. Moreover, inconsistent lighting conditions throughout the recording could cause such disparity together with head movements, which are likely to have an important impact on SCLI in both cases (B.EVM and A.EVM) [1,14].

To the best of our knowledge, our study stands out due to the extensive size of our dataset, which includes a relatively large number of subjects and video recordings (79 video recordings of 65 participants), as well as a relatively wider PR range (measured PR values from 45 bpm to 160 bpm) in comparison to the previously reported results (45 is the largest number of subjects that we found in the literature [61] and the widest PR ranging from 60 bpm to 150 bpm [1,76 ]). Although some studies have included a higher number of video recordings, such as 240 recordings from 21 subjects in [71] and 660 recordings from 33 subjects in [69], our dataset comprises a larger number of unique participants, which can potentially contribute to better variability and generalizability of the results. Our results are comparable to findings published in the study by Renne et al. [1], where participants' PR was extracted from video recordings in the setting of an improvised driving simulator with a low-cost web camera. Renne et al. indicated that there is no visible correlation between the reference BVP and extracted signals from the video recordings, which aligns with our findings. Additionally, the deviations in the detected pulse compared to reference pulse extracted from wearable devices reported in their study are higher than the deviations we present in Figure 5, suggesting that our method yields more accurate results. Our study further demonstrates that the video analysis method can detect statistically significant changes in PR between different age groups of participants in our study, while Renne et al. did not explore the method reliability between different age groups of participants scenarios.

This study is retrospective in nature, as the recordings were originally made for a different purpose [18]. Specifically, the video data were collected to monitor measurement conditions during a driving simulation, rather than to extract PR information from the video. As a result, the recordings include subject movement and non-standard lighting conditions. Additionally, the framing of the shot influences the visibility of key facial regions and the consistency of face positioning, both of which can impact signal extraction. In our case, the dimensions of the extracted ROI are 104 × 104 pixels, which is about 28 times smaller than the dimensions of the smaller original frame (640 × 480). Despite these challenges, our database aligns with the recommendations of previous studies [9,10,77] that emphasize the importance of evaluating pulse extraction methods in dynamic recording environments with higher number of video recordings in variable lighting conditions. This further highlights the need for approaches that are robust to real-world conditions. Given driving simulator conditions, it is expected that signals obtained from the video camera would be sensitive to lighting conditions as well as subject movement. The obtained SNR values for SCLI consistently show higher average values (19.18 dB B.EVM and 17.79 dB A.EVM) compared to the SNR value calculated from the BVP signal (7.67 dB) obtained from the Empatica E4 sensor (Figure 4). The SCLI obtained B.EVM exhibits the highest average SNR value with a standard deviation value (5.42 dB), while the standard deviation of the SNR value when EVM is applied is 4.57 dB. The standard deviation of SNR for BVP signals obtained from the Empatica E4 sensor is 2.18 dB. These values indicate that the data obtained from the Empatica E4 sensor are less noisy due to the lower standard deviation value. Medarević et al. [66] point out that although the SNR values in their work (17.5 ± 3.2 dB at rest) are lower for BVP obtained from Empatica E4 compared to the reference Faros 360 device, they are still sufficiently high to enable reliable state detection. However, the pronounced variability and increased frequency of false positive detections at moderate and low arousal levels indicate the limitations of the Empatica E4 device, primarily due to its sensitivity to motion artifacts, which may cause low SNR in our case as well.

Bent et al. [64] show that deviations in PR recorded using Empatica E4 sensors became more pronounced at higher PR values, such as after physical activity, where Empatica E4 sensors tend to overestimate values compared to standard ECG measurements. Thus, we decide to correct the extracted PR using linear fitting to reduce the possible systematic deviations that potentially exist in the reference data. Furthermore, as stated by Stuyck et al. [65], Empatica E4 often records higher PR values than those registered by ECG devices, which is why Stuyck et al. [65] recommend applying correction factors to reduce sensor deviation. A linear trend of increasing deviation in pulse measurements obtained from the Empatica E4, when compared to the Faros 360 device equipped with high-grade ECG sensors, was also reported by Medarević et al. [66].

Since ECG data from the Faros 360 were not available for our study, pulse measurements recorded by the Empatica E4 were used as reference values. These values, while not laboratory grade, are commonly utilized in research and provide a reasonable basis for comparison [8]. However, these measurements are further corrected using two different linear fit correction parameters (Figure 2) because such a trend is also noticed in the results presented in [66]. One of the reasons for the deviation in the linear trends designed using our data and the data from the Faros 360 and Empatica E4 devices may lie in the fact that the Faros 360 does not measure PPG, but ECG signals. By comparing the results in Figure 5 (top and middle panels), before and after linear fit correction designed using Empatica E4 and video-based PR data (Figure 2, left panel), a decrease in the mean value and standard deviation of MAE and RMSE can be observed. Additionally, the application of linear fit correction leads to a reduction in other MAE and RMSE metrics as well, with AAE decreasing from 10.46 bpm to 6.61 bpm, SAE from 8.83 bpm to 4.54 bpm, and ARE from 0.14 to 0.09 B.EVM. As a result of EVM, the average value of AAE decreases from 10.55 bpm to 5.09 bpm, while SAE decreases from 8.29 bpm to 3.93 bpm and ARE from 0.15 to 0.07. When a linear fit correction, which exists between the Faros 360 and Empatica E4 devices (Figure 2, right panel) [66], is applied to the extracted PR values from the video camera, there is a decrease in MAE and RMSE (Figure 5 top and bottom panels) in both cases (B.EVM and A.EVM). These results indicate that linear fit correction plays an important role in improving pulse detection accuracy, which further supports previous findings suggesting that signals obtained from wearable sensors like the Empatica E4 may exhibit systematic biases or deviations [78].

The average MAE for PR values is 6.48 bpm in the B.EVM case, with a standard deviation of 0.41 bpm. Additionally, the average RMSE value is 7.84 bpm, with a standard deviation of 0.54 bpm (Figure 5, middle panel). It is important to note several key factors contributing to calculated deviations:

1) The conditions under which the video recordings were made are highly non-standardized, with expressed variations in lighting and considerable subject movement during recording, which may introduce impact on PR assessment [1,10,12].
2) Furthermore, the subjects have pronounced individual differences like hairstyles and wearing glasses. These variations further complicate the signal extraction process, as additional elements on the face (like protective medical masks) can alter how skin color is detected via video recordings. [45]

EVM results in low accuracy improvement of the extracted PR values compared to B.EVM results. In the middle panel of Figure 5, it can be seen that the average MAE for PR values after linear fit correction and A.EVM is 5.04 bpm, with a standard deviation of 0.37 bpm. The average RMSE value is 6.38 bpm, with a standard deviation of 0.51 bpm. The obtained results demonstrate an improvement in accuracy for the A.EVM case compared to the B.EVM case. This reduction of deviations suggests that EVM enhances the extraction of physiological signals from video recordings by amplifying subtle color changes associated with blood flow [6,13]. Given the broad age range of the participants and variations in physical characteristics, such as wearing glasses or different haircuts, this is especially important. However, the disadvantage of implementing EVM is that it adds an additional complexity to the method [13], which may limit its usability in real-time [43]. The EVM increases computational complexity in terms of execution time [46,47]. However, parallelization mitigates this overhead. Across all cases, using four cores reduced execution time by approximately 1.5–5.5 times compared to single-core execution (e.g., from ~999 s to ~177 s in the most demanding case) (Table 3), which is in line with [47], demonstrating that parallelization can accelerate similar algorithms by up to 10-fold.

Non-ideal recording conditions and differences in computational approaches probably contributed to the slightly higher deviations in our results compared to previous published studies [43] where deviations between 3 and 7 bpm were observed for quasi-real-time driving. The results presented in Table 1 demonstrate that our proposed approach achieves competitive performance compared to state-of-the-art methods under various conditions. Specifically, our best results in the driving simulator scenario yielded a low MAE of 5.04 bpm, highlighting the applicability of our method in dynamic environment. When we apply proposed approach to public available datasets [61], our method achieved MAE values ranging from 3.52 ± 0.84 bpm to 9.90 ± 8.37 bpm, depending on the dataset quality and acquisition conditions (mainly because participants were squinting in one case and eye detector that we use performed poorer in such conditions [63]). On the same dataset, Bobbia et al. [61] reported RMSE values ranging from 2.39 to 21.22 bpm, while our method achieved an RMSE of 4.33 bpm, demonstrating competitive performance. The best results (MAE = 5.04 bpm and RMSE = 6.38 bpm) presented in our approach are just slightly inferior than results obtained by applying advanced deep learning methods such as 3DCNN [36], which achieved a lower MAE (2.09 bpm) in some cases but at the cost of a higher RMSE (7.30 bpm). The best result among methods that are not based on CNNs was achieved using the CHROME approach (RMSE = 2.39 bpm) as reported in [61]. However, it is important to note that the recordings described in [61] were conducted under well-controlled conditions. For instance, Kwon et al. [10] achieved an average PR estimation error rate of 1.47% using smartphone video recordings, relying on ICA to enhance the signal extracted from the green channel of facial video, which is a much better result than in our study. Similarly, Lamba et al. [72] used a different ROI selection strategy and FFT-based signal processing, yielding an RMSE of 8.35 bpm when focusing solely on the cheeks for PR extraction, being in a similar range with our results. Poh et al. [11] reported averaged RMSE of 4.63 bpm in a setting with minimal movement, like using a laptop (participants were instructed to refrain from making sudden or big movements), which utilized ICA for non-contact PR estimation, but their setup was more complex and less robust to motion. However, we obtain similar RMSE. FDA-approved devices demonstrated deviations of up to ±5 bpm, being an acceptable range for clinical use [79] and closely related to our results after application of linear fit. In our case, when EVM is applied, an average MAE value of 5.09 bpm is obtained (Figure 5, middle panel). Altogether, the results obtained from the Empatica E4 sensor should be interpreted with caution and considered with a pinch of salt, as they may not always provide fully reliable ground truth under all conditions. Several studies [25–27] have reported that Empatica E4, although widely used, can produce less reliable PR measurements under motion or stress conditions due to its sensitivity to noise due to variable sensor–skin contact, which could lead to changes in the intensity of light reaching the sensor. Our future work will include comparison of video-based methods with other more reliable sensors.

### 4.1. Statistical Analysis of Extracted Pulse Rate Between Age Groups

The reference data show a statistically significant difference in PR between the younger and older groups of drivers. Furthermore, the PR extracted from video recordings using the B.EVM and A.EVM methods also confirms this statistically significant difference (Table 2). A clear difference in mean PR between groups is observed: younger participants exhibit a higher mean PR (80.99 bpm) compared to older participants (73.12 bpm), with a difference of approximately 7 bpm. This observed trend is in line with findings reported in the literature. For instance, Umetani et al. [75] and Jose et al. [80] highlighted a general decline in resting heart rate with increasing age in large population samples. However, deviations from this trend may occur under specific conditions. For example, research conducted in driving simulator environments has demonstrated that heart rate can increase in response to elevated cognitive demand or stress, particularly in older adults [78].

The study by Reimer et al. [81] indicates that older participants may exhibit higher heart rates under cognitively demanding tasks compared to younger adults, potentially due to age-related differences in stress response or workload perception. Overall, the heart rate responses are task and context dependent and may not always follow the general trend of age-related decline. Our results indicate that, even after applying linear fit correction, a statistically significant difference remains between the older and younger groups of participants in both cases (B.EVM and A.EVM).

### 4.2. Contributions of the Study

Main contributions in this paper are as follows:

1. The proposed method deals with data originally collected for a completely different purpose, which we made publicly available [82], demonstrating its applicability in non-ideal and real-world scenarios. To the best of our knowledge, this is the first successful application of rPPG in driving simulator as another research group failed to extract the pulse remotely [1].
2. Although there are notable differences between reference and rPPG data (low cross-correlation), the comparison between younger and older drivers reveals meaningful and statistically significant differences, demonstrating that the proposed method can detect the meaningful physiological differences confirmed by the measurements from the reference device, making it an efficient wireless or contactless alternative to different wearable devices currently used in majority of driving simulation studies. Thus, we believe that our manuscript presents interesting results in the area of applied science.
3. We identified a linear trend in the deviation between the PR values estimated from the video recordings and the reference values measured by the Empatica E4 sensor. A similar linear trend was previously reported between the Empatica E4 and the Faros 360 device [66], which supports the consistency and interpretability of our results.
4. Primarily, we would argue that our research opens further exploration of rPPG application in driving simulators.

### 4.3. Limitations of the Study and Future Improvements

We note the following limitations and suggest avenues for further investigation:

1. One of the key improvements is the optimization of measurement and recording conditions. Standardized lighting conditions could enable more consistent results by eliminating variations in light conditions that currently complicate precise signal extraction [9,43]. Additionally, it would be beneficial to provide the responders with the measuring procedure, which instructs them to move as little as possible throughout the test [9], whenever possible. Moreover, a simple calibration could be conducted with the participant’s eyes closed to minimize facial and head movements, allowing the algorithm to focus on stable regions such as the forehead and cheeks while eliminating signal variations caused by eyelid movements and blinking. This calibration could potentially help define the ROI for further analysis.
2. The quality of the camera is another important factor that can affect PR assessment. Using high-resolution cameras with better light sensors can increase the accuracy of detecting skin color changes, allowing for finer color differences that may currently not be detected accurately enough. An IR camera has proven to be highly effective for face detection, with Nijskens et al. [2] achieving a high percentage of frames where the subject's face is accurately detected, although performance diminishes with participant movement. Meanwhile, even lower-resolution cameras (640 × 480) can achieve good accuracy under specific conditions, such as limited pulse range and minimal movement during recording [10,11,83]. In our future research, we plan to examine thoroughly how camera specifications affect the success of PR measurement.

3. Unlike traditional sliding-window or multi-stage approaches that often incur high computational costs, YOLO efficiently processes entire images in a single forward pass, allowing the system to operate on resource-constrained devices or in embedded systems with limited computational power [28]. These characteristics position YOLO as a practical choice for scalable and deployable real-time face detection within video-based physiological monitoring pipelines [84]. Instead of detecting the face and eyes region, the algorithm could be constructed for direct facial skin segmentation [85]. If only the facial skin is segmented, the algorithm can ignore parts of the image that contain hair, glasses, or background, resulting in a more accurate and reliable signal. The standard YOLO model is designed for object detection [84]. To use it for skin segmentation, it would be necessary to adapt or retrain it. This avenue is especially compelling due to the high processing speed of the YOLO architecture (up to 30 Hz [86]).
4. To further assess the robustness of the proposed method, it would be beneficial to increase the number of participants with a wide range of PR values. In addition, it is of utmost importance to investigate how other factors, such as skin tone, affect the reliability of color-based PR detection methods. Previous studies have shown that melanin content should not significantly influence the signal quality in remote photoplethysmography, potentially reducing accuracy in individuals with darker skin tones [87,88].
5. An important step toward enhancing the accuracy and reliability of the method is the use of alternative reference sensors. These sensors could serve as a more accessible and practical reference for evaluating the extracted PR and IBI signals from video recordings.
6. The integration of advanced machine learning algorithms presents a promising avenue for enhancing the detection and prevention of deepfake technologies, which are becoming an increasingly prevalent and sophisticated threat in the field of biometric authentication and security [89]. These algorithms can be specifically tailored to detect subtle physiological cues, such as pulse signals, extracted from video recordings [90]. By analyzing these subtle variations in the skin tone and facial features that are often missed by the human eye, machine learning models can differentiate between genuine biometric data and artificially generated deepfakes [90]. This capability is important not only for safeguarding personal identity verification systems but also for broader applications in cybersecurity, where the accuracy and reliability of biometric data are paramount. Future advancements in this area could lead to more robust security protocols that are resilient against the ever-evolving landscape of deepfake technology, ensuring the integrity of biometric systems in a wide range of applications, from secure access control to forensic investigations.

## 5. Conclusion

This study investigates the feasibility of camera-based pulse rate assessment in a driving simulator, contributing to the development of contactless human–machine interaction. We extract physiological signals from facial recordings and compare them with reference signals obtained from the Empatica E4 wearable sensor. The results indicate that while video-based methods offer a promising alternative to traditional contact sensors, challenges remain in achieving better accuracy and robustness.

The results of the statistical analysis show that the extracted PR from video recordings follow the reference values obtained from the Empatica E4 sensor despite relatively low cross-correlation coefficients in relation to the reference recordings. When a statistically significant difference is detected between older and younger driver groups in the reference data, this difference is also confirmed in the PR extracted from the video, regardless of whether EVM is applied or whether linear fit correction is performed. Al-together, the results indicate that EVM does not greatly impact the quality of pulse assessment in this context, suggesting that the method could be applied in quasi-real-time settings for practical use.

This work has implications for future rPPG-based contactless human-machine interaction within driving simulators, where contactless physiological monitoring could enhance comfort especially in scenarios such as virtual driving training, fatigue detection, or stress-aware user interfaces. We demonstrate the feasibility of applying rPPG under non-ideal, real-world conditions using data originally collected for a different purpose, addressing the lack of available studies on rPPG in driving simulator environments. The method detects meaningful physiological differences between younger and older drivers in alignment with reference device measurements, supporting its potential as a contactless alternative to wearable sensors. Additionally, identifying a consistent linear trend in deviations between video-based and reference PR measurements reinforces the interpretability of our results. Future work will focus on optimizing measurement conditions and employing advanced machine learning techniques to further improve accuracy and robustness.

**Author Contributions**: Conceptualization, N.M.; methodology, Đ.D.N., J.S., and N.M.; software, Đ.D.N.; validation, K.S.P.; formal analysis, Đ.D.N. and N.M.; investigation, K.S.P. and J.S.; data curation, K.S.P. and J.S.; writing—original draft preparation, Đ.D.N.; writing—review and editing, K.S.P., J.S., and N.M.; visualization, Đ.D.N. All authors have read and agreed to the published version of the manuscript.

**Funding:** Nadica Miljković was financially supported by the Ministry of Science, Technological Development and Innovation of the Republic of Serbia under contract No. 451-03-137/2025-03/200103. The work presented in this paper was financially supported by the Slovenian Research Agency within program ICT4QL, grant no. P2-0246 for Kristina Stojmenova Pečečnik and Jaka Sodnik.

**Institutional Review Board Statement:** The study was conducted in accordance with the Declaration of Helsinki, and the protocol was approved by the Institutional Review Board of the Department of ICT (875597) on 1 September 2022.

**Informed Consent Statement:** Informed consent was obtained from all subjects involved in the study. Moreover, written consent was signed by the additional subject, who agreed with the presentation of their facial image in the paper. The subject unequivocally approved the photography representation in Figure 1.

**Data Availability Statement:** All data with identifying parameters (to comply to the subjects' anonymity) are available from the first Author on request. The dataset is shared openly on the Zenodo repository with a Creative Commons Attribution 4.0 International license.

**Acknowledgments:** We express our deep gratitude to Jelena Medarević, from the Faculty of Electrical Engineering, University of Ljubljana for her valuable ideas regarding the analysis of signals from the Empatica E4 sensor and for providing us with additional data for assessing Empatica E4 bias (we cite her paper). During the preparation of this study, the first author utilized GPT3.5 (ChatGPT) to enhance clarity and language and the first author is fully responsible for the publication content after using this tool.

**Conflicts of Interest:** The authors declare that they have no known competing financial interests or personal relationships that could have appeared to influence the work reported in this paper.

## References

1. Renner, P.; Gleichauf, J.; Winkelmann, S. Non-Contact In-Car Monitoring of Heart Rate: Evaluating the Eulerian Video Magnification Algorithm in a Driving Simulator Study. In Proceedings of the Mensch und Computer 2024, Karlsruhe, Germany, 1–4 September 2024; pp. 651–654. https://doi.org/10.1145/3670653.3677493.

2. Nijskens, L.; van der Hurk, S.E.; van den Broek, S.P.; Louvenberg, S.; Souman, J.L.; Bos, J.E.; ter Haar, F.B. An EO/IR monitoring system for noncontact physiological signal analysis in automated vehicles. In Proceedings of the SPIE Autonomous Systems for Security and Defence, Edinburgh, UK, 13 November 2024; Volume 13207, pp. 55–68. https://doi.org/10.1117/12.3033956.
3. Gaur, P.; Temple, D.S.; Hegarty-Craver, M.; Boyce, M.D.; Holt, J.R.; Wenger, M.F.; Preble, E.A.; Eckhoff, R.P.; McCombs, M.S.; Davis-Wilson, H.C.; et al. Continuous Monitoring of Heart Rate Variability in Free-Living Conditions Using Wearable Sensors: Exploratory Observational Study. *JMIR Form. Res.* **2024**, *8*, e53977. https://doi.org/10.2196/53977.
4. Medarević, J.; Tomažič, S.; Sodnik, J. Simulation-based driver scoring and profiling system. *Heliyon* **2024**, *10*, e40310. https://doi.org/10.1016/j.heliyon.2024.e40310.
5. Boboc, R.G.; Butilă, E.V.; Butnariu, S. Leveraging wearable sensors in virtual reality driving simulators: A review of techniques and applications. *Sensors* **2024**, *24*, 4417.
6. Sun, Y.; Thakor, N. Photoplethysmography revisited: From contact to noncontact, from point to imaging. *IEEE Trans. Biomed. Eng.* **2015**, *63*, 463–477. https://doi.org/10.1109/TBME.2015.2476337.
7. Dudarev, V.; Barral, O.; Zhang, C.; Davis, G.; Enns, J.T. On the reliability of wearable technology: A tutorial on measuring heart rate and heart rate variability in the wild. *Sensors* **2023**, *23*, 5863. https://doi.org/10.3390/s23135863.
8. Ronca, V.; Martinez-Levy, A.C.; Vozzi, A.; Giorgi, A.; Aricò, P.; Capotorto, R.; Borghini, G.; Babiloni, F.; Di Flumeri, G. Wearable technologies for electrodermal and cardiac activity measurements: A comparison between fitbit sense, empatica E4 and shimmer GSR3+. *Sensors* **2023**, *23*, 5847. https://doi.org/10.3390/s23135847.
9. Lewandowska, M.; Nowak, J. Measuring pulse rate with a webcam. *J. Med. Imaging Health Inform.* **2012**, *2*, 87–92. https://doi.org/10.1166/jmihi.2012.1064.
10. Kwon, S.; Kim, H.; Park, K.S. Validation of heart rate extraction using video imaging on a built-in camera system of a smartphone. In Proceedings of the 2012 Annual International Conference of the IEEE Engineering in Medicine and Biology Society, San Diego, CA, USA, 28 August–1 September 2012; IEEE: Piscataway, NJ, USA, 2012; pp. 2174–2177. https://doi.org/10.1109/EMBC.2012.6346392.
11. Poh, M.Z.; McDuff, D.J.; Picard, R.W. Non-contact, automated cardiac pulse measurements using video imaging and blind source separation. *Opt. Express* **2010**, *18*, 10762–10774. https://doi.org/10.1364/OE.18.010762.
12. Ernst, H.; Scherpf, M.; Malberg, H.; Schmidt, M. Optimal color channel combination across skin tones for remote heart rate measurement in camera-based photoplethysmography. *Biomed. Signal Process. Control* **2021**, *68*, 102644. https://doi.org/10.1016/j.bspc.2021.102644.
13. Wu, H.Y.; Rubinstein, M.; Shih, E.; Guttag, J.; Durand, F.; Freeman, W. Eulerian video magnification for revealing subtle changes in the world. *ACM Trans. Graph. (TOG)* **2012**, *31*, 1–8. https://doi.org/10.1145/2185520.2185561.
14. Miljković, N.; Trifunović, D. Pulse rate assessment: Eulerian video magnification vs. electrocardiography recordings. In Proceedings of the 12th Symposium on Neural Network Applications in Electrical Engineering (NEUREL), Belgrade, Serbia, 25–27 November 2014; IEEE: Piscataway, NJ, USA, 2012; pp. 17–20. https://doi.org/10.1109/NEUREL.2014.7011447.
15. Wadhwa, N.; Rubinstein, M.; Durand, F.; Freeman, W.T. Phase-based video motion processing. *ACM Trans. Graph. (ToG)* **2013**, *32*, 1–10. https://doi.org/10.1145/2461912.2461966.
16. Balakrishnan, G.; Durand, F.; Guttag, J. Detecting pulse from head motions in video. In Proceedings of the IEEE Conference on Computer Vision and Pattern Recognition, Portland, OR, USA, 23–28 June 2013; pp. 3430–3437. https://doi.org/10.1109/CVPR.2013.440.
17. Lomaliza, J.P.; Park, H. Detecting pulse from head motions using smartphone camera. In Proceedings of the International Conference on Advanced Engineering Theory and Applications, Busan, Vietnam, 8–10 December 2016; Springer International Publishing: Cham, Switzerland, 2016; pp. 243–251. https://doi.org/10.1007/978-3-319-50904-4_24.
18. Gruden, T.; Pececnik, K.S.; Jakus, G.; Sodnik, J. Quantifying Drivers' Physiological Responses to Take-Over Requests in Conditionally Automated Vehicles. In Proceedings of the Human-Computer Interaction Slovenia 2022, Ljubljana, Slovenia, 29 November 2022.

https://doi.org/10.1016/j.aap.2022.106766.
19. Raybaut, P. Spyder-Documentation. 2009. Available online: https://www.spyder-ide.org/ (accessed on 26 August 2025).
20. Harris, C.R.; Millman, K.J.; Van Der Walt, S.J.; Gommers, R.; Virtanen, P.; Cournapeau, D.; Oliphant, T.E. Array programming with NumPy. *Nature* **2020**, *585*, 357–362. https://doi.org/10.1038/s41586-020-2649-2.
21. Bradski, G.; Kaehler, A. OpenCV. Dr. Dobb's J. Softw. Tools 2000, 3. Available online: https://github.com/opencv/opencv/wiki/CiteOpenCV (accessed on 26 August 2025).
22. Virtanen, P.; Gommers, R.; Oliphant, T.E.; Haberland, M.; Reddy, T.; Cournapeau, D.; Burovski, E.; Peterson, P.; Weckesser, W.; Bright, J.; et al. SciPy 1.0: Fundamental algorithms for scientific computing in Python. *Nat. Methods* **2020**, *17*, 261–272. https://doi.org/10.1038/s41592-019-0686-2.
23. Pedregosa, F.; Varoquaux, G.; Gramfort, A.; Michel, V.; Thirion, B.; Grisel, O.; Blondel, M.; Prettenhofer, P.; Weiss, R.; Dubourg, V.; et al. Scikit-learn: Machine learning in Python. *J. Mach. Learn. Res.* **2011**, *12*, 2825–2830.
24. Van Rossum, G. *The Python Library Reference, Release 3.8. 2.*; Python Software Foundation: Beaverton, OR, USA, 2020.
25. McCarthy, C.; Pradhan, N.; Redpath, C.; Adler, A. Validation of the Empatica E4 wristband. In Proceedings of the 2016 IEEE EMBS International Student Conference (ISC), Ottawa, ON, Canada, 29–31 May 2016; IEEE: Piscataway, NJ, USA, 2016; pp. 1–4. https://doi.org/10.1109/EMBSISC.2016.7508621.
26. Schuurmans, A.A.T.; de Looff, P.; Nijhof, K.S.; Rosada, C.; Scholte, R.H.J.; Popma, A.; Otten, R. Validity of the Empatica E4 wristband to measure heart rate variability (HRV) parameters: A comparison to electrocardiography (ECG). *J. Med. Syst.* **2020**, *44*, 1–11. https://doi.org/10.1007/s10916-020-01648-w.
27. Van Voorhees, E.E.; Dennis, P.A.; Watkins, L.L.; Patel, T.A.; Calhoun, P.S.; Dennis, M.F.; Beckham, J.C. Ambulatory heart rate variability monitoring: Comparisons between the empatica e4 wristband and holter electrocardiogram. *Biopsychosoc. Sci. Med.* **2022**, *84*, 210–214. https://doi.org/10.1097/psy.0000000000001010.
28. Jocher, G.; Chaurasia, A.; Qiu, J. YOLOv8 Docs by Ultralytics (Version 8.0. 0). [software]. Available online: https://github.com/ ultralytics/ultralytics (accessed on 26 August 2025).
29. Wang, Y.Q. An analysis of the Viola-Jones face detection algorithm. *Image Process. Line* **2014**, *4*, 128–148. https://doi.org/10.5201/ipol.2014.104.
30. Viola, P.; Jones, M. Rapid object detection using a boosted cascade of simple features. In Proceedings of the 2001 IEEE Computer Society Conference on Computer Vision and Pattern Recognition, CVPR 2001, Kauai, HI, USA, 8–14 December 2001; IEEE: Piscataway, NJ, USA, 2001; Volume 1, pp. 511-518. https://doi.org/10.1109/CVPR.2001.990517.
31. Li, X.; Komulainen, J.; Zhao, G.; Yuen, P.C.; Pietikäinen, M. Generalized face anti-spoofing by detecting pulse from face videos. In Proceedings of the 2016 23rd International Conference on Pattern Recognition (ICPR), Cancun, Mexico, 4–8 December 2016; IEEE: Piscataway, NJ, USA, 2016; pp. 4244–4249. https://doi.org/10.1109/ICPR.2016.7900300.
32. Gonzalez, R.C. *Digital Image Processing*; Pearson Education India: Delhi, India, 2009.
33. Kasinski, A.; Schmidt, A. The architecture and performance of the face and eyes detection system based on the Haar cascade classifiers. *Pattern Anal. Appl.* **2010**, *13*, 197–211. https://doi.org/10.1007/s10044-009-0150-5.
34. Rudinskaya, E.; Paringer, R. *Face Detection Accuracy Study Based on Race and Gender Factor Using Haar Cascades*; CEUR Workshop Proceedings: Aachen, Germany, 2020; Volume 2667, pp. 238–242.
35. Yu, S.G.; Kim, S.E.; Kim, N.H.; Suh, K.H.; Lee, E.C. Pulse rate variability analysis using remote photoplethysmography signals. *Sensors* **2021**, *21*, 6241. https://doi.org/10.3390/s21186241.
36. Speth, J.; Vance, N.; Flynn, P.; Bowyer, K.; Czajka, A. Unifying frame rate and temporal dilations for improved remote pulse detection. *Comput. Vis. Image Underst.* **2021**, *210*, 103246. https://doi.org/10.1016/j.cviu.2021.103246.
37. Lu, L.; Zhu, T.; Morelli, D.; Creagh, A.; Liu, Z.; Yang, J.; Rullan, A.; Clifton, L.; Pimentel, M.A.F.; Tarassenko, L.; et al. Uncertainties in the analysis of heart rate variability: A systematic review.

IEEE Rev. Biomed. Eng. **2023**, *17*, 180–196. https://doi.org/10.1109/RBME.2023.3271595.
38. Clifford, G.; Sameni, R.; Ward, J.; Robinson, J.; Wolfberg, A.J. Clinically accurate fetal ECG parameters acquired from maternal abdominal sensors. *Am. J. Obstet. Gynecol.* **2011**, *205*, 47.e1–47.e5. https://doi.org/10.1016/j.ajog.2011.02.066.
39. Developed with the Special Contribution of the European Heart Rhythm Association (EHRA); Endorsed by the European Association for Cardio-Thoracic Surgery (EACTS); Authors/Task Force Members; Camm, A.J.; Kirchhof, P.; Lip, G.Y.H.; Schotten, U.; Savelieva, I.; Ernst, S.; Van Gelder, I.C.; et al. Guidelines for the management of atrial fibrillation: The Task Force for the Management of Atrial Fibrillation of the European Society of Cardiology (ESC). *Eur. Heart J.* **2010**, *31*, 2369–2429. https://doi.org/10.1093/eurheartj/ehq278.
40. Nussinovitch, U.; Elishkevitz, K.P.; Kaminer, K.; Nussinovitch, M.; Segev, S.; Volovitz, B.; Nussinovitch, N. The efficiency of 10-second resting heart rate for the evaluation of short-term heart rate variability indices. *Pacing Clin. Electrophysiol.* **2011**, *34*, 1498–1502. https://doi.org/10.1111/j.1540-8159.2011.03178.x.
41. Tanasković, I.; Miljković, N. A new algorithm for fetal heart rate detection: Fractional order calculus approach. *Med. Eng. Phys.* **2023**, *118*, 104007. https://doi.org/10.1016/j.medengphy.2023.104007.
42. Zhang, Q.; Wu, Q.; Zhou, Y.; Wu, X.; Ou, Y.; Zhou, H. Webcam-based, non-contact, real-time measurement for the physiological parameters of drivers. *Measurement* **2017**, *100*, 311–321. https://doi.org/10.1016/j.measurement.2017.01.007.
43. Hussain, Y.; Shkara, A.A. Speed up Eulerian Video Motion Magnification. *Kurd. J. Appl. Res.* **2017**, *2*, 14–17. https://doi.org/10.24017/science.2017.3.14.
44. Klabunde, R. *Cardiovascular Physiology Concepts*; Lippincott Williams & Wilkins: Philadelphia, PA, USA, 2011.
45. Wang, C.; Pun, T.; Chanel, G. A comparative survey of methods for remote heart rate detection from frontal face videos. *Front. Bioeng. Biotechnol.* **2018**, *6*, 33. https://doi.org/10.3389/fbioe.2018.00033.
46. Lim, K.S.; Moya-Bello, E.; Chavarria-Zamora, L. Resource Optimization of the Eulerian Video Magnification Algorithm Towards an Embedded Architecture. In Proceedings of the 2021 IEEE URUCON, Montevideo, Uruguay, 24–26 November 2021; IEEE: Piscataway, NJ, USA, 2021; pp. 576–579. https://doi.org/10.1109/URUCON53396.2021.9647386.
47. Zhang, K.; Jin, X.; Wu, A. Accelerating Eulerian video magnification using FPGA. In Proceedings of the 2017 19th International Conference on Advanced Communication Technology (ICACT), PyeongChang, Republic of Korea, 19–22 February 2017; IEEE: Piscataway, NJ, USA, 2017; pp. 554–559. https://doi.org/10.23919/ICACT.2017.7890151.
48. Bishop, C.M.; Nasrabadi, N.M. *Pattern Recognition and Machine Learning*; Springer: New York, NY, USA, 2006; Volume 4, Number 4, p. 738.
49. Huang, R.; Hong, K.S.; Yang, D.; Huang, G. Motion artifacts removal and evaluation techniques for functional near-infrared spectroscopy signals: A review. *Front. Neurosci.* **2022**, *16*, 878750. https://doi.org/10.3389/fnins.2022.878750.
50. Sathyapriya, L.; Murali, L.; Manigandan, T. Analysis and detection R-peak detection using Modified Pan-Tompkins algorithm. In Proceedings of the 2014 IEEE International Conference on Advanced Communications, Control and Computing Technologies, Ramanathapuram, India, 8–10 May 2014; IEEE: Piscataway, NJ, USA, 2014; pp. 483–487. https://doi.org/10.1109/ICACCCT.2014.7019490.
51. Nayak, C.; Saha, S.K.; Kar, R.; Mandal, D. An optimally designed digital differentiator based preprocessor for R-peak detection in electrocardiogram signal. *Biomed. Signal Process. Control.* **2019**, *49*, 440–464. https://doi.org/10.1016/j.bspc.2018.09.005.
52. Johnson, M.J.; Chahal, T.; Stinchcombe, A.; Mullen, N.; Weaver, B.; Bédard, M. Physiological responses to simulated and on-road driving. *Int. J. Psychophysiol.* **2011**, *81*, 203–208. https://doi.org/10.1016/j.ijpsycho.2011.06.012.
53. Kohlhaas, M.; Seidlmayer, L.; Kaspar, M. A Specialized System for Arrhythmia Detection for Basic Research in Cardiology. In *German Medical Data Sciences: Bringing Data to Life*; IOS Press: Amsterdam, The Netherlands, 2021; pp. 3–7. https://doi.org/10.3233/shti210041.
54. Rumaling, M.I.; Chee, F.P.; Bade, A.; Goh, L.P.W.; Juhim, F. Biofingerprint detection of corona

virus using Raman spectroscopy: A novel approach. *SN Appl. Sci.* **2023**, *5*, 197. https://doi.org/10.1007/s42452-023-05419-3.
55. Hauke, J.; Kossowski, T. Comparison of values of Pearson's and Spearman's correlation coefficients on the same sets of data. *Quaest. Geogr.* **2011**, *30*, 87–93. https://doi.org/10.2478/v10117-011-0021-1.
56. Smith, L.I. *A Tutorial on Principal Components Analysis*; University of Otago: Otago, New Zealand, 2002.
57. Yu, Z.; Li, X.; Zhao, G. Remote photoplethysmograph signal measurement from facial videos using spatio-temporal networks. *arXiv* **2019**, arXiv:1905.02419. https://doi.org/10.48550/arXiv.1905.02419.
58. Garbarino, M.; Lai, M.; Bender, D.; Picard, R.W.; Tognetti, S. Empatica E3—A wearable wireless multi-sensor device for real-time computerized biofeedback and data acquisition. In Proceedings of the 2014 4th International Conference on Wireless Mobile Communication and Healthcare-Transforming Healthcare Through Innovations in Mobile and Wireless Technologies (MOBIHEALTH), Athens, Greece, 3–5 November 2014; IEEE: Piscataway, NJ, USA, 2014; pp. 39–42. http://dx.doi.org/10.4108/icst.mobihealth.2014.257418.
59. Biswas, D.; Simões-Capela, N.; Van Hoof, C.; Van Helleputte, N. Heart rate estimation from wrist-worn photoplethysmography: A review. *IEEE Sens. J.* **2019**, *19*, 6560–6570. https://doi.org/10.1109/JSEN.2019.2914166.
60. Xiao, H.; Liu, T.; Sun, Y.; Li, Y.; Zhao, S.; Avolio, A. Remote photoplethysmography for heart rate measurement: A review. *Biomed. Signal Process. Control.* **2024**, *88*, 105608. https://doi.org/10.1016/j.bspc.2023.105608.
61. Bobbia, S.; Macwan, R.; Benezeth, Y.; Mansouri, A.; Dubois, J. Unsupervised skin tissue segmentation for remote photoplethysmography. *Pattern Recognit. Lett.* **2019**, *124*, 82–90. https://doi.org/10.1016/j.patrec.2017.10.017.
62. Benezeth, Y.; Krishnamoorthy, D.; Monsalve, D.J.B.; Nakamura, K.; Gomez, R.; Mitéran, J. Video-based heart rate estimation from challenging scenarios using synthetic video generation. *Biomed. Signal Process. Control.* **2024**, *96*, 106598. https://doi.org/10.1016/j.bspc.2024.106598.
63. Kamarudin, N.; Jumadi, N.A.; Mun, N.L.; Keat, N.C.; Ching, A.H.K.; Mahmud, W.M.H.W.; Morsin, M.; Mahmud, F. Implementation of haar cascade classifier and eye aspect ratio for driver drowsiness detection using raspberry Pi. *Universal J. Electr. Electron. Eng.* **2019**, *6*, 67–75. https://doi.org/10.13189/ujeee.2019.061609.
64. Bent, B.; Goldstein, B.A.; Kibbe, W.A.; Dunn, J.P. Investigating sources of inaccuracy in wearable optical heart rate sensors. *NPJ Digit. Med.* **2020**, *3*, 18. https://doi.org/10.1038/s41746-020-0226-6.
65. Stuyck, H.; Dalla Costa, L.; Cleeremans, A.; Van den Bussche, E. Validity of the Empatica E4 wristband to estimate resting-state heart rate variability in a lab-based context. *Int. J. Psychophysiol.* **2022**, *182*, 105–118. https://doi.org/10.1016/j.ijpsycho.2022.10.003.
66. Medarević, J.; Miljković, N.; Stojmenova Pečečnik, K.; Sodnik, J. Distress Detection in VR environment using Empatica E4 wristband and Bittium Faros 360. *Front. Physiol.* **2025**, *16*, 1480018. https://doi.org/10.3389/fphys.2025.1480018.
67. Hartikainen, S.; Lipponen, J.A.; Hiltunen, P.; Rissanen, T.T.; Kolk, I.; Tarvainen, M.P.; Martikainen, T.J.; Castrén, M.; Väliaho, E.-S.; Jäntti, H. Effectiveness of the chest strap electrocardiogram to detect atrial fibrillation. *Am. J. Cardiol.* **2019**, *123*, 1643–1648. https://doi.org/10.1016/j.amjcard.2019.02.028.
68. Hess, M.R.; Kromrey, J.D. Robust confidence intervals for effect sizes: A comparative study of Cohen's d and Cliff's delta under non-normality and heterogeneous variances. In Proceedings of the Annual Meeting of the American Educational Research Association, San Diego, CA, USA, 12–16 April 2004; Volume 1.
69. Tang, J.; Chen, K.; Wang, Y.; Shi, Y.; Patel, S.; McDuff, D.; Liu, X. Mmpd: Multi-domain mobile video physiology dataset. In Proceedings of the 2023 45th Annual International Conference of the IEEE Engineering in Medicine & Biology Society (EMBC), Sydney, Australia, 24–27 July 2023; IEEE: Piscataway, NJ, USA, 2023; pp. 1–5. https://doi.org/10.1109/EMBC40787.2023.10340857.
70. Lee, H.; Cho, A.; Whang, M. Fusion method to estimate heart rate from facial videos based on RPPG and RBCG. *Sensors* **2021**, *21*, 6764. https://doi.org/10.3390/s21206764.

71. Ma, X.; Tang, J.; Jiang, Z.; Cheng, S.; Shi, Y.; Li, D.; Zhang, T.; Liu, H.; Chen, L.; Zhao, Q.; et al. Non-Contact Health Monitoring During Daily Personal Care Routines. *arXiv* **2025**, arXiv:2506.09718. https://doi.org/10.48550/arXiv.2506.09718.
72. Lamba, P.S.; Virmani, D. Contactless heart rate estimation from face videos. *J. Stat. Manag. Syst.* **2020**, *23*, 1275–1284. https://doi.org/10.1080/09720510.2020.1799584.
73. Lee, H.; Ko, H.; Chung, H.; Nam, Y.; Hong, S.; Lee, J. Real-time realizable mobile imaging photoplethysmography. *Sci. Rep.* **2022**, *12*, 7141. https://doi.org/10.1038/s41598-022-11265-x.
74. Fallet, S.; Schoenenberger, Y.; Martin, L.; Braun, F.; Moser, V.; Vesin, J.M. Imaging photoplethysmography: A real-time signal quality index. In Proceedings of the 2017 Computing in Cardiology (CinC), Rennes, France, 24–27 September 2017; IEEE: Piscataway, NJ, USA, 2017; pp. 1–4. https://doi.org/10.22489/CinC.2017.142-139.
75. Umetani, K.; Singer, D.H.; McCraty, R.; Atkinson, M. Twenty-four hour time domain heart rate variability and heart rate: Relations to age and gender over nine decades. *J. Am. Coll. Cardiol.* **1998**, *31*, 593–601. https://doi.org/10.1016/S0735-1097(97)00554-8.
76. Ruba, M.; Jeyakumar, V.; Gurucharan, M.K.; Kousika, V.; Viveka, S. Non-contact pulse rate measurement using facial videos. In Proceedings of the 2020 IEEE International Conference on Advances and Developments in Electrical and Electronics Engineering (ICADEE), Coimbatore, India, 10–11 December 2020; IEEE: Piscataway, NJ, USA, 2020; pp. 1–6. https://doi.org/10.1109/ICADEE51157.2020.9368944.
77. Fukunishi, M.; Kurita, K.; Yamamoto, S.; Tsumura, N. Non-contact video-based estimation of heart rate variability spectrogram from hemoglobin composition. *Artif. Life Robot.* **2017**, *22*, 457–463. https://doi.org/10.1007/s10015-017-0382-1.
78. Ravindran, K.K.; Della Monica, C.; Atzori, G.; Lambert, D.; Revell, V.; Dijk, D.J. Evaluating the Empatica E4 derived heart rate and heart rate variability measures in older men and women. In Proceedings of the 2022 44th Annual International Conference of the IEEE Engineering in Medicine & Biology Society (EMBC), Glasgow, UK, 11–15 July 2022; IEEE: Piscataway, NJ, USA, 2022; pp. 3370–3373. https://doi.org/10.1109/EMBC48229.2022.9871559.
79. Lee, C.; Lee, C.; Fernando, C.; Chow, C.M. Comparison of Apple watch vs KardiaMobile: A tale of two devices. *CJC Open* **2022**, *4*, 939–945. https://doi.org/10.1016/j.cjco.2022.07.011.
80. Jose, A.D.; Collison, D. The normal range and determinants of the intrinsic heart rate in man. *Cardiovasc. Res.* **1970**, *4*, 160–167. https://doi.org/10.1093/cvr/4.2.160.
81. Reimer, B.; Mehler, B.L.; Pohlmeyer, A.E.; Coughlin, J.F.; Dusek, J.A. The use of heart rate in a driving simulator as an indicator of age-related differences in driver workload. Adv. Transp. Stud. Int. J. 2006, 9–20. Available online: https://www.atsinternationaljournal.com/2006-issues/the-use-of-heart-rate-in-a-driving-simulator-as-an-indicator-ofage-related-differences-in-driver-workload/ (accessed on 26 August 2025).
82. Nešković, Đ.D.; Stojmenova Pečečnik, K.; Sodnik, J.; Miljković, N. Dataset comprising extracted R, G, and B components for assessment of remote photopletismography (Version 1) [Data set]. *Zenodo* **2025**. https://doi.org/10.5281/zenodo.16414188.
83. Suh, K.H.; Lee, E.C. Contactless physiological signals extraction based on skin color magnification. *J. Electron. Imaging* **2017**, *26*, 063003. https://doi.org/10.1117/1.JEI.26.6.063003.
84. Garg, D.; Goel, P.; Pandya, S.; Ganatra, A.; Kotecha, K. A deep learning approach for face detection using YOLO. In Proceedings of the 2018 IEEE Punecon, Pune, India, 30 November–2 December 2018; IEEE: Piscataway, NJ, USA, 2018; pp. 1–4. https://doi.org/10.1109/PUNECON.2018.8745376.
85. Phung, S.L.; Bouzerdoum, A.; Chai, D. Skin segmentation using color pixel classification: Analysis and comparison. *IEEE Trans. Pattern Anal. Mach. Intell.* **2005**, *27*, 148–154. https://doi.org/10.1109/TPAMI.2005.17.
86. Redmon, J.; Divvala, S.; Girshick, R.; Farhadi, A. You only look once: Unified, real-time object detection. In Proceedings of the IEEE Conference on Computer Vision and Pattern Recognition, Las Vegas, NV, USA, 27–30 June 2016; pp. 779–788. https://doi.org/10.1109/CVPR.2016.91.
87. Talukdar, D.; De Deus, L.F.; Sehgal, N. The Evaluation of Remote Monitoring Technology Across Participants with Different Skin Tones. *Cureus* **2023**, *15*, e45075. https://doi.org/10.7759/cureus.45075.
88. Talukdar, D.; de Deus, L.F.; Sehgal, N. Evaluation of Remote Monitoring Technology across

different skin tone participants. *MedRxiv* **2023**, https://doi.org/10.1101/2023.04.02.23288057.
89. Das, R.; Negi, G.; Smeaton, A.F. Detecting deepfake videos using euler video magnification. *arXiv* **2021**, arXiv:2101.11563. https://doi.org/10.48550/arXiv.2101.11563.
90. Hernandez-Ortega, J.; Tolosana, R.; Fierrez, J.; Morales, A. Deepfakes detection based on heart rate estimation: Single-and multi-frame. In *Handbook of Digital Face Manipulation and Detection: From DeepFakes to Morphing Attacks*; Springer International Publishing: Cham, Switzerland, 2022; pp. 255–273. https://doi.org/10.1007/978-3-030-87664-7_12.